\documentclass[11pt]{article}

\usepackage[a4paper,text={15.25cm,22cm},centering]{geometry}
\usepackage{amsmath}  
\usepackage{amsfonts} 
\usepackage{graphicx} 

\usepackage{amssymb}
\usepackage{subfigure}
\usepackage{color}

\providecommand\bnabla{\boldsymbol{\nabla}}
\providecommand\bcdot{\boldsymbol{\cdot}}

\newcommand{\be}[1]{\begin{equation}\label{#1}}
\newcommand{\ee}{\end{equation}}
\newcommand{\ba}[1]{\begin{eqnarray}\label{#1}}
\newcommand{\ea}{\end{eqnarray}}
\newcommand{\rf}[1]{(\ref{#1})}
\newcommand{\nn}{\nonumber}

\begin{document}
\title{Singular diffusionless limits of double-diffusive instabilities in magnetohydrodynamics}
\author{Oleg N. Kirillov $^{1,2}$}
\date{\small $^1$Northumbria University, Newcastle upon Tyne, NE1 8ST, UK, E-mail: oleg.kirillov@northumbria.ac.uk
\\
$^2$Steklov Mathematical Institute, Russian Academy of Sciences,
Gubkina 8, Moscow 119991, Russia, E-mail: kirillov@mi.ras.ru}

\maketitle

\begin{abstract}
We study local instabilities of a differentially rotating viscous flow of electrically conducting incompressible fluid subject to an external azimuthal magnetic field.
In the presence of the magnetic field the hydrodynamically stable flow can demonstrate non - axisymmetric azimuthal magnetorotational instability (AMRI) both in the diffusionless case and in the double-diffusive case with viscous and ohmic dissipation. Performing stability analysis of amplitude transport equations of short-wavelength approximation, we find that the threshold of the diffusionless AMRI via the Hamilton-Hopf bifurcation is a singular limit of the thresholds of the viscous and resistive AMRI corresponding to the dissipative Hopf bifurcation and manifests itself as the Whitney umbrella singular point. A smooth transition between the two types of instabilities is possible only if the magnetic Prandtl number is equal to unity, $\rm Pm=1$.
At a fixed $\rm Pm\ne 1$ the threshold of  the double-diffusive AMRI is displaced by finite distance in the parameter space with respect to the diffusionless case even in the zero dissipation limit.
The complete neutral stability surface contains three Whitney umbrella singular points and two mutually orthogonal intervals of self-intersection. At these singularities the double-diffusive system reduces to a marginally stable system which is either Hamiltonian or parity-time (PT) symmetric.
\end{abstract}

\maketitle

\section{Introduction}

While common sense tends to assign to dissipation the role of a vibration damper,
as early as 1879 Kelvin and Tait predicted viscosity-driven instability of Maclaurin's spheroids (proven by Roberts and Stewartson in 1963 \cite{BO2014,C1984,RS1963}) thus presenting a class of Hamiltonian equilibria, which, although stable in the absence of dissipation, become unstable due to the action of dissipative forces \cite{BKMR94,KM2007}. The universality of the \textit{dissipation-induced instabilities} manifests itself in unexpected links between solid- and fluid mechanics \cite{BD2007,KV10,K2013dg}. For instance, the destabilizing action of viscous dissipation on the negative energy mode of rotation of a particle moving in a rotating cavity \cite{Lamb1908} selects backward whirling in the rotating frame as an unstable (anticyclonic) motion. Remarkably, this very instability mechanism described by Lamb in 1908 has recently re-appeared as a trigger breaking the cyclone-anticyclone vortex symmetry in a rotating fluid in the presence of linear Ekman friction \cite{Chefranov2016}.

The onset of the classical Hopf bifurcation in a near-Hamiltonian dissipative system
generically does not converge to the onset of the Hamilton-Hopf bifurcation of a Hamiltonian system when dissipation tends to zero \cite{L2003}. For instance, the onset of secular instability (classical Hopf) of viscous Maclaurin's spheroids does not tend to the onset of dynamical instability (Hamilton-Hopf) of inviscid Maclaurin's spheroids in the limit of vanishing viscosity \cite{RS1963,C1984,BO2014}.
In meteorology this phenomenon is known as the ``Holop\"ainen instability mechanism'' for a baroclinic flow when
waves that are linearly stable in the absence of Ekman friction become dissipatively destabilized in its presence, with the result that the location of the curve of marginal stability is displaced by an order one distance in the parameter space, even if the Ekman number is infinitesimally small \cite{H1961,KM2007,S2010,WE2012}. A similar effect in solid mechanics is represented by the ``Ziegler destabilization paradox'' \cite{Ziegler52,KV10,B1956,KS2005,T2016}.

Swaters noticed in \cite{S2010} that the stability boundary
associated with the zero dissipation limit of a dissipative baroclinic instability theory does
not collapse to the inviscid result when the Ekman dissipation is replaced by other dissipative mechanisms, e.g. by horizontal turbulent friction, confirming that such a \textit{singular limit} is generic. However, he also managed to choose a specific dissipative perturbation (in which the dissipation is proportional to the geostrophic potential vorticity) possessing coincidence of the zero dissipation limit of the dissipative marginal stability boundary with the inviscid result \cite{S2010}.

The destabilization by dissipation is especially intriguing when several diffusion mechanisms act simultaneously \cite{Yih1961,T1974,LD1977,C1984,ORT1986,M1993}. In this case ``no simple rule for the effect of introducing small viscosity or diffusivity on flows that are neutral in their absence appears to hold'' \cite{TSL2013}. In hydrodynamics, a classical example is given by secular instability of the Maclaurin spheroids due to both fluid viscosity and gravitational radiation reaction, where the critical eccentricity of the meridional section of the spheroid depends on the \textit{ratio} of the two dissipative mechanisms and reaches its maximum, corresponding to the onset of dynamical instability in the ideal system, exactly at this ratio equal to 1 \cite{LD1977,C1984}.
In solid mechanics, the generic character of the discontinuity of the instability threshold in the zero dissipation limit was already noticed  in the work by Smith \cite{S1933,K2009} who found that a viscoelastic shaft rotating in bearings with viscous damping is prone to dissipation-induced instability for almost all ratios of the damping coefficient of the shaft and the damping coefficient of the bearings, except one specific ratio.

In hydrodynamics and MHD the ratio of damping coefficients corresponding to different dissipative mechanisms is traditionally called \textit{the Prandtl number}. For example, the Prandtl number, ${\rm Pr}=\nu/\kappa$, measures the relative strength of the diffusion of vorticity represented in the Navier-Stokes equations by the kinematic viscosity coefficient $\nu$ and thermal diffusion with the coefficient of thermal diffusivity $\kappa$ \cite{AG1978}. The magnetic Prandtl number, ${\rm Pm}=\nu/{\eta}$, is the ratio of the coefficients of the kinematic viscosity and ohmic diffusion, $\eta$ \cite{BH2008pm,B2011,LK1985}. To get an idea of the key role of the Prandtl numbers in the correspondence between stability criteria in the diffusionless and the double-diffusive case let us consider the Rayleigh centrifugal instability criterion and its extensions.

The Rayleigh criterion predicts a stationary axisymmetric instability of an ideal incompressible Newtonian fluid, differentially rotating with the radially-varying angular velocity $\Omega=\Omega(r)$, if
\be{irc}
{\rm Ro}+1<0,
\ee
where $\rm Ro$ is the fluid Rossby number
\be{hro}
{\rm Ro}:=\frac{r\partial_r\Omega}{2\Omega}
\ee
and $\partial_r=\frac{\partial}{\partial r}$.
For a viscous fluid the Rayleigh criterion \rf{irc} is modified as follows \cite{EY95}
\be{vrc}
{\rm Ro}+1+\frac{1}{4{\rm Re}^2}<0
\ee
and reduces to the diffusionless criterion \rf{irc} as the Reynolds number, ${\rm Re}\rightarrow \infty$.

In the general multiple-diffusive case the existence of such a direct correspondence between the diffusionless and diffusive stability criteria is not evident. In many cases, however, the reduction of the double-diffusive instability criteria to the diffusionless ones can be achieved by setting the corresponding Prandtl number to a specific value, e.g. to 1, and then tending diffusivities to zero (or, equivalently, the corresponding Reynolds numbers to infinity) \cite{LD1977}.

For example, the stationary axisymmetric instability known as the double-diffusive Goldreich-Schubert-Fricke (GSF) instability \cite{AG1978,MMLC2013} develops in a rotating viscous and thermally conducting  fluid when the extended Rayleigh criterion is fulfilled \cite{AG1978}
\be{gsf1}
4({\rm Ro}+1)+{\rm Pr}\frac{N^2}{\Omega^2}+\frac{1}{{\rm Re}^2}<0,
\ee
where $N$ is the Brunt-V\"ais\"al\"a frequency\footnote{
which in the limit of $\gamma \rightarrow \infty$ reduces to the buoyancy frequency in the Boussinesq approximation
$N^2=-\frac{g}{\rho}\frac{d \rho}{d r}$.} \cite{BP2016}
$$
N^2:=\frac{g}{\gamma}\frac{\partial}{\partial r}\ln{(p \rho^{-{\gamma}})}=
\frac{g}{\gamma}\left(\frac{1}{p}\frac{\partial p}{\partial r}-\frac{\gamma}{\rho}\frac{\partial \rho}{\partial r}\right),
$$
and $p$ is the pressure of the fluid, $\rho$ the density, $\gamma$ the adiabatic index, and $g$ the radial acceleration. When dissipative effects are absent, $\nu=0$, $\kappa=0$,
the diffusionless GSF instability occurs for \cite{AG1978}
\be{gsf2}
4({\rm Ro}+1)+\frac{N^2}{\Omega^2}<0.
\ee
Evidently, ${\rm Pr}=1$ is the only value at which the criterion \rf{gsf1} reduces to \rf{gsf2} in the limit ${\rm Re}\rightarrow \infty$.

Similarly, Michael's criterion  of ideal  MHD \cite{M1954}
predicts stationary axisymmetric instability caused by an azimuthal magnetic field for a rotating flow of a non-viscous incompressible Newtonian fluid that is a perfect electrical conductor, if\cite{M1954}
\be{micr}
{\rm Ro}+1-\frac{\omega_{A_{\phi}}^2}{\Omega^2}{\rm Rb}<0,
\ee
where $\rm Rb$ is the magnetic Rossby number \cite{KS13}
\be{mro}
{\rm Rb}:=\frac{r\partial_r\omega_{A_{\phi}}}{2\omega_{A_{\phi}}}
\ee
and $\omega_{A_{\phi}}$ is the $\rm Alfv\acute{e}n$ angular velocity related to the magnitude of the magnetic field \cite{OP1996}.
Again, the diffusionless Michael's criterion \rf{micr} follows in the limit of ${\rm Re}\rightarrow \infty$ from its double-diffusive counterpart\footnote{By analysing the criterion \rf{micrdd} for the case of the rigid-body rotation ($\rm Ro =0$), Acheson and Hide \cite{A1973} came to the conclusion that the electrical resistance of the fluid opposes the destabilizing influence of a radial increase of the magnetic field (when ${\rm Rb}>0$) and that viscous effects support it. They supposed that ``the damping effect of viscosity on the disturbances is offset by its action as an agent for diffusion of momentum, which reduces the stabilizing effect of rigid-body rotation by enabling the circulation of a displaced ring of fluid to harmonize more readily with its surroundings.'' These ideas were further developed in \cite{AG1978,A1980,P1981} with application to double-diffusive convection in a stratified fluid.} \cite{AG1978,KSF2014b}
\be{micrdd}
{\rm Ro}+1-{\rm Pm}\frac{\omega_{A_{\phi}}^2}{\Omega^2}{\rm Rb}+\frac{1}{4{\rm Re}^2}<0
\ee
only if ${\rm Pm}=1$.

In particular, Michael's criterion for both the diffusionless and the double-diffusive problem  predicts stability with respect to  axisymmetric perturbations for the rotating flow and the azimuthal magnetic field that satisfy the following constraints
\be{ces}
\Omega=\omega_{A_{\phi}},\quad {\rm Ro}={\rm Rb}=-1.
\ee

In 1956 Chandrasekhar \cite{Chandra56} observed that the properties \rf{ces} correspond to
an exact steady solution of the MHD equations for an incompressible fluid in the ideal case, i.e. when $\nu=0$ and $\eta=0$. For this solution the total pressure of the fluid and the magnetic field is constant, the fluid velocity at every point is parallel to the direction of the magnetic field at that point, and the $\rm Alfv\acute{e}n$ angular velocity is equal to the angular velocity of the fluid, which implies
equality of the densities of the fluid magnetic and kinetic energies. This {\it energy equipartition solution} of the ideal MHD was proven by Chandrasekhar \cite{Chandra56} to be marginally stable against \textit{general} perturbations. \footnote{A bit surprisingly, as he admitted in his memoirs \cite{CH2010}: ``One nice result which nevertheless came out at this time was the proof of the stability of the equipartition solution. Wentzel and Goldberger checked my analysis as I could not quite believe the result myself.'' Actually, the Chandrasekhar equipartition solution belongs to a wide class of exact stationary solutions of MHD equations for the case of ideal incompressible infinitely conducting fluid with constant total pressure that includes even flows with knotted magnetic surfaces \cite{GK2012}.}

To illustrate stability of the equipartition solution  \rf{ces} with respect to non-axisymmetric perturbations, we
substitute it into the following criterion of destabilization of  a hydrodynamically stable rotating flow of an inviscid and perfectly conducting fluid by an azimuthal magnetic field
\be{opbh}
\frac{\omega_{A_{\phi}}^2}{\Omega^2}<-\frac{4{\rm Ro}}{m^2},
\ee
where $m\gg 1$ is the azimuthal wavenumber and ${\rm Ro}<0$ \cite{AG1978,BH1992,OP1996}. The criterion \rf{opbh} is valid in the limit of infinitely large axial and azimuthal  wavenumbers of the perturbation. Naturally, the solution \rf{ces} violates \rf{opbh} already at $m\ge2$, thus confirming the Chandrasekhar theorem \cite{Chandra56}.

Recently, Bogoyavlenskij \cite{Bog2004} discovered that viscous and resistive incompressible MHD equations
possess exact \textit{unsteady} equipartition solutions with finite and equal kinetic and magnetic energies  when
the fluid velocity and the magnetic field are collinear and the kinematic viscosity $\nu$ is equal to the magnetic diffusivity $\eta$, i.e. when ${\rm Pm}=1$. Under the constraint ${\rm Pm}=1$ the Bogoyavlenskij unsteady equipartition solutions turn into the ideal and \textit{steady} Chandrasekhar equipartition equilibria when $\nu=\eta\rightarrow 0$ \cite{Bog2004}.


One could expect that in double-diffusive MHD the remarkable stability of the Chandrasekhar energy equipartition solution is preserved under the constraint ${\rm Pm}=1$. As soon as the constraint is violated, one could anticipate a dissipation-induced instability of the  equipartition solution. For instance, recent analytical works \cite{KS13,KSF2014b} demonstrated that in the inductionless limit\footnote{A very small ratio of viscosity of the fluid to its electrical resistivity, typically of order $10^{-6}-10^{-5}$, is a characteristic of liquid metals that are used in laboratory experiments, e.g., with the magnetized Couette-Taylor flow \cite{S2014} and von Karman flow \cite{VKS2007}. Recently developed Newtonian ferrofluids have a tunable $\rm Pm$ in the diapason from $10^{-6}$ to $10^{-4}$ \cite{Carle2017}. In astrophysics and geophysics such small values of $\rm Pm$ are typical for the planetary interiors, cold parts of accretion disks, and ``dead-zones'' of the protoplanetary disks \cite{BP2016,JB2013,KS13,KSF2014b,CKH2015,RM2015,B2011,BH2008pm,LK1985}.} of ${\rm Pm}= 0$  a rotating viscous incompressible fluid with vanishing electrical conductivity is destabilized by azimuthal magnetic fields of arbitrary radial dependency, if
\be{rorb}
8({\rm Ro}+1){\rm Rb}>-({\rm Ro}+2)^2.
\ee
The inequality \rf{rorb} predicts the onset of the azimuthal magnetorotational instability (AMRI) even in the case of the Keplerian rotating flow with $\rm Ro=-3/4$ when $\rm Rb>-25/32$ \cite{KS13}.
In particular, \rf{rorb} implies destabilization of the Chandrasekhar equipartition solution, whose susceptibility to the double-diffusive AMRI at ${\rm Pm}\ll 1$ has been confirmed numerically in \cite{CKH2015,RM2015}.

According to the group-theoretical argument by Julien and Knobloch \cite{JK2010} AMRI is an oscillatory instability with a nonzero azimuthal wavenumber, which is most likely to develop in the presence of the azimuthal magnetic filed \cite{S2014,KSF2012}. Hence, its onset in the double-diffusion case is characterized by the classical Hopf bifurcation, at which simple eigenvalues cross the imaginary axis in the complex plane. On the other hand, the equations of the diffusionless MHD can be written in Hamiltonian form \cite{M1990}. For this reason, the stable oscillatory nonaxisymmetric modes in the ideal MHD case can carry both positive and negative energy; their interaction yields the Hamilton-Hopf bifurcation at the onset of the non-axisymmetric oscillatory instabilities \cite{IKS2009}.

In the present work we perform a local stability analysis of a circular Couette-Taylor flow of a viscous and electrically conducting fluid in an azimuthal magnetic field of arbitrary radial dependence. We obtain a unifying geometric picture that naturally connects the diffusionless and double-diffusive AMRI in low- and high-${\rm Pm}$ regimes in the spirit of the singularity theory approach by
Bottema \cite{B1956}, Arnold \cite{A1971}, and Langford \cite{L2003} on generic singularities in the multiparameter families of matrices, which is especially efficient when
combined with the perturbation of multiple eigenvalues, index theory and exploitation of the fundamental symmetries of the ideal system \cite{M1991,MO1995,BD2007,K2013pt,K2013dg}.

After a brief re-derivation of the already known WKB equations of the system we write the corresponding algebraic eigenvalue problem, which determines the dispersion relation, as a non-Hamiltonian perturbation of a Hamiltonian eigenvalue problem. The latter yields the dispersion relation of the ideal system. This allows us to investigate systematically the singular limit of the onset of the oscillatory AMRI due to the classical Hopf bifurcation at arbitrary $\rm Pm$ when viscous and resistive terms tend to  zero.

In the frame of the local stability analysis we show that the threshold of the double-diffusive AMRI tends to the threshold of the diffusionless AMRI only at ${\rm Pm}=1$ as the Reynolds numbers tend to infinity and find the Whitney umbrella singularity on the neutral stability surface that dictates this specific choice of $\rm Pm$.
 We classify the stable oscillatory modes involved in the Hamilton-Hopf bifurcation by their Krein (or energy) sign. Then, we explicitly demonstrate by means of the perturbation theory for eigenvalues that when viscosity and ohmic diffusivity are weak (and even infinitesimally small), the dominance of viscosity destroys stability of the negative energy mode at ${\rm Pm}>1$ whereas the dominance of ohmic diffusivity destabilizes the positive energy mode at ${\rm Pm}<1$ (including the inductionless case ${\rm Pm}=0$) in the close vicinity of the Hamilton-Hopf bifurcation. However, when the fluid Rossby number exceeds some critical value, the destabilization is possible only at finite values of Reynolds numbers and is accompanied by a
transfer of instability between negative- and positive-energy modes that occurs due to the presence of complex exceptional points in the spectrum. This clarifies the reasons for instability of Chandrasekhar's equipartition solution and its extensions at both low and high $\rm Pm$.

\section{Transport equation for amplitudes and its dispersion relation}
\subsection{Governing equations and the background fields}
The dynamics of a flow of a viscous and electrically conducting incompressible fluid that interacts with the magnetic field is described by
the Navier-Stokes equation for the fluid velocity $\boldsymbol{u}$ which is coupled with the induction equation for the magnetic field $\boldsymbol{B}$ \cite{BP2016,KSF2014b}
\ba{m1}
&\frac{\partial \boldsymbol{u}}{\partial t}+\boldsymbol{u} \bcdot \bnabla \boldsymbol{u}-\frac{1}{\mu_0 \rho}\boldsymbol{B}\bcdot \bnabla\boldsymbol{B} +\frac{1}{\rho} \bnabla P-\nu \bnabla^2 \boldsymbol{u}=0,&\nn \\
&\frac{\partial \boldsymbol{B}}{\partial t}+\boldsymbol{u} \bcdot \bnabla \boldsymbol{B} - \boldsymbol{B} \bcdot \bnabla \boldsymbol{u}- \eta \bnabla^2 \boldsymbol{B}=0.&
\ea
In the equations \rf{m1} the total pressure is denoted by $P=p+\frac{\boldsymbol{B}^2}{2\mu_0}$, $p$ is the hydrodynamic pressure, $\rho=const$
the density, $\nu=const$ the kinematic viscosity, $\eta=(\mu_0 \sigma)^{-1}$ the magnetic diffusivity, $\sigma=const$ the conductivity of the fluid,
and $\mu_0$ the magnetic permeability of free space. In addition, the incompressible flow and the solenoidal magnetic field fulfil the constraints:
\be{m3}
\bnabla \bcdot \boldsymbol{u} = 0,\quad  \bnabla \bcdot \boldsymbol{B}=0.
\ee

It is well-known that for a flow differentially rotating in a gap between the radii
$r_1$ and $r_2>r_1$ the equations \rf{m1} and \rf{m3} possess a steady solution of the general form  \cite{CKH2015,S2009}
\be{m4} \boldsymbol{u}_0(r)=r\,\Omega(r)\,\boldsymbol{
e}_{\phi},\quad p=p_0(r), \quad \boldsymbol{B}_0(r)=B_{\phi}^0(r)\boldsymbol{
e}_{\phi}
\ee
in the cylindrical coordinate system $(r, \phi, z)$.
In the \textit{magnetized circular Couette-Taylor flow} \rf{m4} the angular velocity profile $\Omega(r)$ and the
azimuthal magnetic field $B_{\phi}^0(r)$ are arbitrary functions of the radial coordinate $r$ satisfying boundary conditions for an inviscid and non-resistive fluid \cite{CKH2015,S2009}.  For a viscous and resistive fluid, the angular velocity has the form $\Omega(r)=a+b r^{-2}$, while the expression for the magnetic
field is given by $B_{\phi}^0(r)=c r + d r^{-1}$ with the coefficients determined from boundary conditions \cite{CKH2015,S2009}. In the following we will perform a local linear stability analysis of the magnetized circular Couette-Taylor flow for a viscous and resistive fluid.

In 1956 Chandrasekhar \cite{Chandra56} observed that for the exact stationary solution \rf{m4} of the equations \rf{m1} and \rf{m3} with $\Omega=\frac{B_{\phi}^0}{r\sqrt{\rho\mu_0}}$ and $P=const$  in the ideal case, i.e. when $\nu=0$ and $\eta=0$, the kinetic and magnetic energies are in equipartition, $\frac{\rho(\Omega r)^2}{2}=\frac{(B_{\phi}^0)^2}{2\mu_0}$, and ${\rm Ro}={\rm Rb}=-1$. The latter equality follows from the condition of constant total pressure and from the fact that in the steady-state the centrifugal acceleration of the background flow is compensated by the pressure gradient, $r\Omega^2=\frac{1}{\rho}\partial_r p_0$ \cite{KSF2014b}.
Note that ${\rm Ro}=-1$ corresponds to the velocity profile $\Omega(r)\sim r^{-2}$ whereas ${\rm Rb}=-1$ corresponds to the magnetic field produced by an axial current $I$ isolated from the fluid \cite{KSF2014b,S2014,CKH2015}: $B_{\phi}^0(r)=\frac{\mu_0 I}{2\pi r}$.

Linearizing equations \rf{m1}-\rf{m3} in the vicinity of the stationary solution \rf{m4} by assuming general perturbations $\boldsymbol{ u}=\boldsymbol{ u}_0+\boldsymbol{ u}'$, $p=p_0+p'$, and $\boldsymbol{ B}=\boldsymbol{ B}_0+\boldsymbol{ B}'$, leaving only the terms of first order with respect to the primed quantities, and introducing the gradients of the background fields represented by the two $3 \times 3$ matrices
\ba{L3a}
&\mathcal{U}:=\bnabla \boldsymbol{ u}_0=\Omega\left(
                                \begin{array}{ccc}
                                  0 & -1 & 0 \\
                                  1+2{\rm Ro} & 0 & 0 \\
                                  0 & 0 & 0 \\
                                \end{array}
                              \right), \quad
\mathcal{B}:=\bnabla \boldsymbol{ B}_0=\frac{B_{\phi}^0}{r}\left(
 \begin{array}{ccc}
                                                                              0 & -1 & 0 \\
                                                                              1+2{\rm Rb} & 0 & 0 \\
                                                                              0 & 0 & 0 \\
                                                                            \end{array}
                                                                          \right),&
\ea
we arrive at the linearized system of magnetohydrodynamics  \cite{KSF2014b,KK2013,KS13}
\ba{L4}
\left(
  \begin{array}{cc}
    \partial_t+\mathcal{U}+\boldsymbol{ u}_0 \bcdot \bnabla-\nu \bnabla^2 & -\frac{\mathcal{B}+\boldsymbol{ B}_0 \bcdot \bnabla}{\rho\mu_0} \\
    \mathcal{B}-
\boldsymbol{B}_0\bcdot \bnabla & \partial_t-\mathcal{U}+\boldsymbol{ u}_0 \bcdot \bnabla-\eta \bnabla^2 \\
  \end{array}
\right)\left(
             \begin{array}{c}
               \boldsymbol{ u}'  \\
               \boldsymbol{ B}'  \\
             \end{array}
           \right)=
           -\frac{\bnabla}{\rho}\left(
                     \begin{array}{c}
                        p'+
\frac{\boldsymbol{ B}_0 \bcdot \boldsymbol{ B}'}{ \mu_0} \\
                       0 \\
                     \end{array}
                   \right),
\ea
where the perturbations fulfil the constraints
\be{L2}
\bnabla \bcdot \boldsymbol{ u}' = 0,\quad  \bnabla \bcdot \boldsymbol{ B}'=0.
\ee

\subsection{Derivation of the amplitude transport equations}

Let $\epsilon$ be a small parameter ($0 <\epsilon \ll 1$). We
seek solutions of the linearized equations \rf{L4} in the form of asymptotic expansions
with respect to the small parameter $\epsilon$ \cite{E1987}:
\ba{g1}
\boldsymbol{ L}'(\boldsymbol{ x},t,\epsilon)=e^{i\Phi(\boldsymbol{ x},t)/\epsilon}\left(\boldsymbol{ L}^{(0)}(\boldsymbol{ x},t)+\epsilon\boldsymbol{ L}^{(1)}(\boldsymbol{ x},t,) \right)+\epsilon \boldsymbol{L}^{(r)}(\boldsymbol{ x},t,\epsilon),
\ea
where $\boldsymbol{L}'=(\boldsymbol{ u}',\boldsymbol{ B}',p')^T$, $\boldsymbol{L}^{(j)}=(\boldsymbol{ u}^{(j)},\boldsymbol{ B}^{(j)},p^{(j)})^T$,  $\boldsymbol{ x}$ is a vector of coordinates, $\Phi$ represents the phase of the wave or the eikonal, and  $\boldsymbol{ u}^{(j)}$, $\boldsymbol{ B}^{(j)}$, and $p^{(j)}$, $j=0,1,r$, are complex-valued amplitudes. The index $r$ denotes the remainder terms that are assumed to be uniformly bounded in $\epsilon$ on any fixed time interval \cite{L1991,LSB1996}.

Maslov \cite{Maslov1986} observed that high-frequency oscillations $\exp(i\epsilon^{-1}\Phi(\boldsymbol{ x},t))$ quickly
die out because of viscosity unless one assumes a quadratic dependency of
viscosity on the small parameter $\epsilon$. Following \cite{Maslov1986,AS2015,EY95}, we assume
that $\nu=\epsilon^2\widetilde \nu$ and $\eta=\epsilon^2\widetilde \eta$.

Substituting expansions \rf{g1} in \rf{L4} and
collecting terms at $\epsilon^{-1}$ and $\epsilon^0$,
we find \cite{KSF2014b}
\ba{g3}
&\epsilon^{-1}:\,\left(
  \begin{array}{rr}
    \partial_t \Phi +(\boldsymbol{ u}_0 \bcdot \bnabla \Phi) & -
\frac{(\boldsymbol{ B}_0\bcdot \bnabla\Phi)}{\rho\mu_0} \\
    -(\boldsymbol{ B}_0 \bcdot \bnabla \Phi) & \partial_t \Phi +(\boldsymbol{ u}_0 \bcdot \bnabla \Phi) \\
  \end{array}
\right)\left(
         \begin{array}{c}
           \boldsymbol{ u}^{(0)} \\
           \boldsymbol{ B}^{(0)} \\
         \end{array}
       \right)=-\frac{\bnabla \Phi}{\rho}\left(
                 \begin{array}{c}
                     p^{(0)}+\frac{\boldsymbol{ B}_0\bcdot \boldsymbol{ B}^{(0)} }{\mu_0}\\
                   0 \\
                 \end{array}
               \right)&\nn\\
\ea
\ba{g3a}
&\epsilon^{0}:\,i\left(
  \begin{array}{rr}
    \partial_t \Phi +(\boldsymbol{ u}_0 \bcdot \bnabla \Phi) & -
\frac{(\boldsymbol{ B}_0\bcdot \bnabla\Phi)}{\rho\mu_0} \\
    -(\boldsymbol{ B}_0 \bcdot \bnabla \Phi) & \partial_t \Phi +(\boldsymbol{ u}_0 \bcdot \bnabla \Phi) \\
  \end{array}
\right)\left(
         \begin{array}{c}
           \boldsymbol{ u}^{(1)} \\
           \boldsymbol{ B}^{(1)} \\
         \end{array}
       \right)+i\frac{\bnabla \Phi}{\rho}\left(
                 \begin{array}{c}
                     p^{(1)}+\frac{\boldsymbol{ B}_0\bcdot \boldsymbol{ B}^{(1)}}{\mu_0} \\
                   0 \\
                 \end{array}
               \right)&\nn\\
&+\left(
  \begin{array}{cc}
    \partial_t+\mathcal{U}+\boldsymbol{ u}_0 \bcdot \bnabla+ \widetilde \nu(\bnabla \Phi)^2 & -\frac{\mathcal{B}+\boldsymbol{ B}_0 \bcdot \bnabla}{\rho\mu_0} \\
    \mathcal{B}-
\boldsymbol{B}_0\bcdot \bnabla & \partial_t-\mathcal{U}+\boldsymbol{ u}_0 \bcdot \bnabla+\widetilde \eta(\bnabla \Phi)^2 \\
  \end{array}
\right)\left(
         \begin{array}{c}
           \boldsymbol{ u}^{(0)} \\
           \boldsymbol{ B}^{(0)} \\
         \end{array}
       \right)&\nn\\
&+\frac{\bnabla}{\rho}\left(
                     \begin{array}{c}
                        p^{(0)}+\frac{\boldsymbol{ B}_0\bcdot \boldsymbol{ B}^{(0)}}{\mu_0} \\
                       0 \\
                     \end{array}
                   \right)=0.&
\ea
The solenoidality conditions \rf{L2} yield
\ba{g4}
&\boldsymbol{ u}^{(0)}\bcdot \bnabla\Phi=0,\quad \bnabla\bcdot \boldsymbol{ u}^{(0)}+i\boldsymbol{ u}^{(1)}\bcdot \bnabla \Phi
=0,&\nn\\
&\boldsymbol{ B}^{(0)}\bcdot \bnabla\Phi=0,\quad \bnabla\bcdot \boldsymbol{ B}^{(0)}+i\boldsymbol{ B}^{(1)}\bcdot \bnabla \Phi
=0.&
\ea

Taking the dot product of the first of the equations in the system \rf{g3} with $\bnabla \Phi$ under the constraints \rf{g4} we find that for $\bnabla \Phi \ne 0$
\be{pressa}
p^{(0)}  =-\frac{\boldsymbol{ B}_0\bcdot \boldsymbol{ B}^{(0)}}{\mu_0}.
\ee
Under the condition \rf{pressa} the equation \rf{g3} has a nontrivial solution if the determinant of the $6\times6$ matrix in its left-hand side vanishes. This gives us two characteristic roots corresponding to the two $\rm Alfv\acute{e}n$ waves \cite{KK2013,E1987} that yield the following two Hamilton-Jacobi equations:
\be{g10}
\partial_t \Phi+\left(\boldsymbol{ u}_0\pm\frac{\boldsymbol{ B}_0}{\sqrt{\rho \mu_0}}\right) \bcdot \bnabla\Phi=0.
\ee
The characteristic roots $\left(-\boldsymbol{ u}_0\pm\frac{\boldsymbol{ B}_0}{\sqrt{\rho \mu_0}}\right) \bcdot \bnabla\Phi$ are triple and semi-simple
and degenerate into a semi-simple characteristic root of multiplicity 6 on the surface \cite{KK2013,E1987}
\be{g6a}
\boldsymbol{ B}_0\bcdot \bnabla\Phi=0.
\ee

When \rf{g6a} is fulfilled, the derivative of the phase along the fluid stream lines vanishes:
\be{g6}
\frac{D \Phi}{Dt}:=\partial_t \Phi + \boldsymbol{ u}_0 \bcdot \bnabla \Phi=0.
\ee
Using the relations \rf{pressa}, \rf{g6a}, \rf{g6} we simplify the equations \rf{g3a}:
\ba{g7}
&\left(\frac{D}{Dt}+\widetilde \nu(\bnabla \Phi)^2 +\mathcal{U} \right) \boldsymbol{ u}^{(0)}
-
\frac{1}{\rho\mu_0}(\mathcal{B}+\boldsymbol{ B}_0\bcdot \bnabla)\boldsymbol{ B}^{(0)} =
-\frac{i}{\rho} \left(p^{(1)}+\frac{1}{\mu_0}(\boldsymbol{ B}_0 \bcdot \boldsymbol{ B}^{(1)})\right)\bnabla \Phi,&\nn\\
&\left(\frac{D}{Dt}+\widetilde \eta(\bnabla \Phi)^2 -\mathcal{U} \right) \boldsymbol{B}^{(0)}+(\mathcal{B}-\boldsymbol{ B}_0\bcdot \bnabla)
\boldsymbol{ u}^{(0)}=0.&
\ea

Eliminating pressure in the first of Eqs.~\rf{g7} via multiplication of it by $\bnabla \Phi$ and taking into account the constraints \rf{g4}, then using the identities
\ba{g11}
&\bnabla \partial_t \Phi+\bnabla (\boldsymbol{ u}_0 \bcdot \bnabla)\Phi=\frac{D}{D t}\bnabla \Phi+\mathcal{U}^T\bnabla\Phi=0,&\nn\\
&\bnabla (\boldsymbol{ B}_0 \bcdot \bnabla\Phi)=(\boldsymbol{ B}_0 \bcdot \bnabla) \bnabla \Phi+\mathcal{B}^T\bnabla \Phi=0,&\nn\\
&
\frac{D}{Dt}( \bnabla \Phi\bcdot\boldsymbol{ u}^{(0)})=\frac{D \bnabla \Phi}{Dt}\bcdot\boldsymbol{ u}^{(0)}+\bnabla \Phi\bcdot\frac{D \boldsymbol{ u}^{(0)}}{Dt}=0,&\nn\\
&
(\boldsymbol{ B}_0 \bcdot \bnabla)(\bnabla\Phi \bcdot \boldsymbol{ B}^{(0)})=
((\boldsymbol{ B}_0 \bcdot \bnabla)\bnabla\Phi) \bcdot \boldsymbol{ B}^{(0)}+\bnabla\Phi \bcdot (\boldsymbol{ B}_0 \bcdot \bnabla)\boldsymbol{ B}^{(0)}=0,&
\ea
and, finally,
denoting $\boldsymbol{ k}=\bnabla \Phi$, we write the transport equations for the amplitudes \rf{g7} as
\ba{g13}
\frac{D\boldsymbol{ u}^{(0)}}{Dt}&=&-\left(\mathcal{I}-\frac{2\boldsymbol{ k}\boldsymbol{ k}^T}{|\boldsymbol{ k}|^2}\right)\mathcal{U} \boldsymbol{ u}^{(0)}-\widetilde \nu
|\boldsymbol{ k}|^2\boldsymbol{ u}^{(0)}
+
\frac{1}{\rho\mu_0}
\left(\left(\mathcal{I}-\frac{2\boldsymbol{ k}\boldsymbol{ k}^T}{|\boldsymbol{ k}|^2}\right)\mathcal{B}+\boldsymbol{B}_0\bcdot \bnabla\right)\boldsymbol{ B}^{(0)}, \nn\\
\frac{D \boldsymbol{ B}^{(0)}}{Dt}&=&~~~~~~~ \mathcal{U} \boldsymbol{ B}^{(0)}-\widetilde \eta|\boldsymbol{ k}|^2\boldsymbol{ B}^{(0)}-
(\mathcal{B}-\boldsymbol{ B}_0\bcdot \bnabla) \boldsymbol{ u}^{(0)},
\ea
where $\mathcal{I}$ is the $3\times 3$ identity matrix. From the phase equation \rf{g11} we deduce that
\be{g12}
\frac{D\boldsymbol{ k}}{D t}=-\mathcal{U}^T\boldsymbol{ k}.
\ee
The equations \rf{g12} and \rf{g13} are valid under the assumption that the condition \rf{g6a} is fulfilled.

The local partial differential equations \rf{g13} are fully equivalent to the transport equations of \cite{K2013dg,KSF2014b}. In the case of the ideal MHD when viscosity and resistivity are zero, the equations \rf{g13} exactly coincide with those of the work \cite{KK2013} and are fully equivalent to the transport equations derived in \cite{FV95}. In the absence of the magnetic field these equations are reduced to that of the work \cite{EY95} that considered stability of the viscous Couette-Taylor flow.

Note that the leading order terms dominate the solution \rf{g1} for a sufficiently long time, provided that $\epsilon$ is small enough \cite{L1991,LSB1996}, which reduces analysis of instabilities to the investigation of the growth rates of solutions of the transport equations \rf{g13}.

According to \cite{EY95} and \cite{FV95}, in order to study physically relevant and potentially unstable modes we have to choose bounded and asymptotically non-decaying solutions of the system \rf{g12}. These correspond to $k_{\phi}\equiv 0$ and $k_R$ and $k_z$ time-independent. Note that this solution is
compatible with the constraint $\boldsymbol{ B}_0\bcdot \boldsymbol{ k}=0$ following from \rf{g6a}.

\subsection{Dispersion relation of the double-diffusive amplitude equations}

Denote $\alpha=k_z|\boldsymbol{ k}|^{-1}$, $|\boldsymbol{ k} |^2=k_R^2+k_z^2$ and introduce the $\rm Alfv\acute{e}n$ angular velocity, the viscous and resistive frequencies, and the hydrodynamic and magnetic Reynolds numbers \cite{KSF2014b}:
\be{freq}
\omega_{A_{\phi}}=\frac{B_{\phi}^0}{R\sqrt{\rho\mu_0}},\quad \omega_{\nu}=\widetilde{\nu} |\boldsymbol{ k}|^2,\quad \omega_{\eta}=\widetilde{\eta}|\boldsymbol{ k}|^2,\quad {\rm Re}=\frac{\alpha \Omega}{\omega_{\nu}},\quad {\rm Rm}=\frac{\alpha \Omega}{\omega_{\eta}}.
\ee
In particular, ${\rm Rm}={\rm Re}{\rm Pm}$.

Looking
for a solution to Eqs.~\rf{g13} in the modal form \cite{FV95}: $\boldsymbol{ u}^{(0)}=\widehat{\boldsymbol{ u}} e^{\alpha\Omega\lambda t +im\phi}$, $\boldsymbol{ B}^{(0)}=\sqrt{\rho\mu_0}\widehat{\boldsymbol{ B}} e^{\alpha\Omega\lambda t  +im\phi}$
we write the amplitude equations in the matrix form
\be{ep1}
{\bf A}{\bf z}=\lambda{\bf z},
\ee
where ${\bf z}=(\widehat{u}_R,\widehat{u}_{\phi},\widehat{B}_R,\widehat{B}_{\phi})^T\in\mathbb{C}^4$ and ${\bf A}={\bf A}_0+{\bf A}_1\in\mathbb{C}^{4\times4}$ with \cite{KSF2014b,KS13,SB2014}
\be{mri11}
{\bf A}_0=\left(
    \begin{array}{cccc}
      -in & 2\alpha & i  n {\rm S} & -2\alpha{\rm S}\\
      -\frac{2(1+ {\rm Ro})}{\alpha} & -in & \frac{2(1+{\rm Rb})}{\alpha}{\rm S} & in {\rm S}  \\
      in {\rm S}  & 0 & -in & 0 \\
      -\frac{2{\rm Rb}}{\alpha}{\rm S} & in{\rm S} & \frac{2 {\rm Ro}}{\alpha} & -in \\
    \end{array}
  \right),\quad
{\bf A}_1=\left(
    \begin{array}{cccc}
      \frac{-1}{\rm Re} & 0 & 0 & 0\\
      0 & \frac{-1}{\rm Re} & 0 & 0  \\
      0 & 0 & \frac{-1}{\rm Rm} & 0 \\
      0 & 0 & 0 & \frac{-1}{\rm Rm} \\
    \end{array}
  \right).
\ee
The ratio $n=\frac{m}{\alpha}$ is the modified azimuthal wavenumber and
$
{\rm S}=\frac{\omega_{A_{\phi}}}{\Omega}
$ is the $\rm Alfv\acute{e}n$ angular velocity in the units of $\Omega$.

Let us introduce a Hermitian matrix
\be{mG}
{\bf G=\left(
         \begin{array}{cccc}
           0 & -i & 0 & i{\rm S} \\
           i & 0 & -i{\rm S} & 0 \\
           0 & i{\rm S} & 4\frac{{\rm Ro}-{\rm Rb}}{\alpha n} & -i \\
           -i{\rm S} & 0 & i & 0 \\
         \end{array}
       \right)
}
\ee
and define an indefinite inner product in $\mathbb{C}^4$ as
$
[{\bf x},{\bf y}]=\overline{\bf y}^T{\bf G}{\bf x}
$
\cite{YS1975,K2013dg} and a standard inner product as $({\bf x},{\bf y})=\overline{\bf y}^T{\bf x}$.
The matrix ${\bf H}_0=-i{\bf G}{\bf A}_0$ is Hermitian too:
\be{Hm}
{\bf H}_0=\left(
            \begin{array}{cccc}
              -\frac{2 ({\rm S}^2 {\rm Rb}-{\rm Ro}-1)}{\alpha} & i n ({\rm S}^2+1) & -\frac{2 {\rm S} (1+{\rm Rb}-{\rm Ro})}{\alpha} & -2 i n {\rm S} \\
              -i n ({\rm S}^2+1) & 2 \alpha & 2 i n {\rm S} & -2 \alpha {\rm S}\\
              -\frac{2 {\rm S} (1+{\rm Rb}-{\rm Ro})}{\alpha} & -2 i n {\rm S} & \frac{2 ({\rm S}^2 {\rm Rb}+{\rm S}^2+2 {\rm Rb}-3 {\rm Ro})}{\alpha} & i n ({\rm S}^2+1) \\
              2 i n {\rm S} & -2 \alpha {\rm S} & -i n ({\rm S}^2+1) & 2 \alpha {\rm S}^2  \\
            \end{array}
          \right).
\ee

Consequently, the eigenvalue problem
$
{\bf A}_0{\bf z}=\lambda {\bf z}
$
can be written in the Hamiltonian form with the Hamiltonian ${\bf H}_0$ \cite{YS1975,K2013dg,Z2016}:
\be{hs}
{\bf H}_0{\bf z}=i^{-1}{\bf G}\lambda {\bf z}.
\ee

The fundamental symmetry
\be{fs}
{\bf A}_0=-{\bf G}^{-1}\overline{{\bf A}_0}^T{\bf G},
\ee
where the overbar denotes complex conjugation, implies the symmetry of the spectrum of the matrix ${\bf A}_0$ with respect to the imaginary axis \cite{YS1975,K2013dg}.

The full eigenvalue problem \rf{ep1} is thus a dissipative perturbation of the Hamiltonian eigenvalue problem \rf{hs}
\be{fep}
({\bf H}_0+{\bf H}_1){\bf z}=i^{-1}{\bf G}\lambda {\bf z},
\ee
where ${\bf H}_1=-i{\bf G}{\bf A}_1$ is a complex non-Hermitian matrix:
\be{Dm}
{\bf H}_1=\left(
  \begin{array}{cccc}
    0 & \frac{1}{\rm Re} & 0 & -\frac{\rm S}{\rm Rm} \\
    -\frac{1}{\rm Re} & 0 & \frac{\rm S}{\rm Rm} & 0 \\
    0 & -\frac{\rm S}{\rm Re} & 4i\frac{{\rm Ro}-{\rm Rb}}{\alpha n {\rm Rm}} & \frac{1}{\rm Rm} \\
    \frac{\rm S}{\rm Re} & 0 & -\frac{1}{\rm Rm} & 0 \\
  \end{array}
\right).
\ee
The complex characteristic equation $p(\lambda):=\det({\bf H}_0+{\bf H}_1-i^{-1}{\bf G}\lambda {\bf I})=0$,
where $\bf I$ is the $4\times4$ identity matrix,
is the dispersion relation for the double-diffusive system \rf{fep}.

\section{Linear Hamilton-Hopf bifurcation and the diffusionless AMRI}
\subsection{Krein sign and splitting of double eigenvalues with Jordan block}

Consider the unperturbed (Hamiltonian) case corresponding to ${\bf H}_1=0$.
A simple imaginary eigenvalue $\lambda=i\omega$ of the eigenvalue problem \rf{hs}
with the eigenvector $\bf z$ is said to have positive Krein sign if $[{\bf z}, {\bf z}]>0$
and negative Krein sign if $[{\bf z}, {\bf z}]<0$ \cite{YS1975,K2013dg}.

Denote by $\bf p$ the vector of all parameters of the matrix ${\bf H}_0$: ${\bf p}=({\rm S},{\rm Ro},{\rm Rb},n)^T\in \mathbb{R}^4$. Let at ${\bf p}={\bf p}_0$ the matrix ${\bf H}_0={\bf H}({\bf p}_0)$ have a double imaginary eigenvalue $\lambda=i \omega_0$ $(\omega_0\ge 0)$ with the Jordan chain consisting of the eigenvector ${\bf z}_0$ and the associated vector ${\bf z}_1$ that satisfy the following equations \cite{YS1975,K2013dg}:
\be{ham1}
{\bf H}_0{\bf z}_0=\omega_0{\bf G}{\bf z}_0,\quad {\bf H}_0{\bf z}_1=\omega_0{\bf G}{\bf z}_1+i^{-1}{\bf G}{\bf z}_0.
\ee
Transposing these equations and applying the complex conjugation yields
\be{ham2}
\overline{\bf z}_0^T{\bf H}_0=\omega_0\overline{\bf z}_0^T{\bf G},\quad
\overline{\bf z}_1^T{\bf H}_0=\omega_0\overline{\bf z}_1^T{\bf G}-i^{-1}\overline {\bf z}_0^T {\bf G}.
\ee
As a consequence, $\overline{\bf z}_0^T{\bf G}{\bf z}_0=0$ and $\overline{\bf z}_1^T{\bf G}{\bf z}_0+\overline{\bf z}_0^T{\bf G}{\bf z}_1=0$ or, in the other notation,
\be{ham3}
[{\bf z}_0,{\bf z}_0]=0,\quad [{\bf z}_0,{\bf z}_1]=-[{\bf z}_1,{\bf z}_0].
\ee

Varying parameters along a curve ${\bf p}={\bf p}(\varepsilon)$ $({\bf p}(0)={\bf p}_0)$, where $\varepsilon$ is a real parameter, and assuming the Newton-Puiseux expansions for the double eigenvalue $i\omega_0$ and its eigenvector in powers of $\varepsilon^{1/2}$ when $|\varepsilon|$ is small, we find \cite{K2013dg}
\be{ham4}
\lambda_{\pm}=i\omega_0\pm i\omega_1\varepsilon^{1/2} +o(\varepsilon^{1/2}),\quad {\bf z}_{\pm}={\bf z}_0\pm i\omega_1 {\bf z}_1\varepsilon^{1/2} +o(\varepsilon^{1/2}),
\ee
with
\be{ham5}
\omega_1=\sqrt{i\frac{\overline{\bf z}_0^T\Delta{\bf H}{\bf z}_0}{\overline{\bf z}_1^T{\bf G}{\bf z}_0}},\quad
\Delta {\bf H}=\left. \sum_{s=1}^4\frac{\partial {\bf H}}{\partial p_s}\frac{d p_s}{d \varepsilon}\right|_{\varepsilon=0}=\overline{(\Delta {\bf H})}^T.
\ee
Taking into account that $\overline{\bf z}_0^T\Delta{\bf H}{\bf z}_0$ is real and $\overline{\bf z}_1^T{\bf G}{\bf z}_0$ is imaginary, we assume that $\omega_1>0$, which is a reasonable assumption in view of the fact that $\omega_0>0$ and $|\varepsilon|$ is small. Then, for $\varepsilon>0$, the double eigenvalue $i\omega_0$ splits into two pure imaginary ones $\lambda_{\pm}=i\omega_0 \pm i\omega_1 \sqrt{\varepsilon}$ (stability). When $\varepsilon < 0$, the splitting yields a pair of complex eigenvalues with real parts of different sign (instability). Therefore, varying parameters along a curve ${\bf p}(\varepsilon)$ we have a linear Hamilton-Hopf bifurcation at the point ${\bf p}_0$, which is a regular point of the boundary between the domains of stability and oscillatory instability. The path ${\bf p}(\varepsilon)$ crosses the stability boundary at the point ${\bf p}_0$.

Calculating the indefinite inner product for the perturbed eigenvectors ${\bf z}_{\pm}$ at $\varepsilon>0$, we find \cite{K2013dg}
\be{ham6}
[{\bf z}_{+},{\bf z}_{+}]=+2i\omega_1 \overline{\bf z}_0^T {\bf G}{\bf z}_1 \varepsilon^{{1}/{2}}+o(\varepsilon^{{1}/{2}}), \quad
[{\bf z}_{-},{\bf z}_{-}]=-2i\omega_1 \overline{\bf z}_0^T {\bf G}{\bf z}_1 \varepsilon^{{1}/{2}}+o(\varepsilon^{{1}/{2}}).
\ee
Therefore, the simple imaginary eigenvalue $\lambda_+$ with the eigenvector ${\bf u}_+$ has the Krein sign which is opposite to the Krein sign of the eigenvalue $\lambda_-$ with the eigenvector ${\bf u}_-$. With decreasing $\varepsilon >0$, the imaginary eigenvalues $\lambda_+$ and $\lambda_-$ with opposite Krein signs move towards each other along the imaginary axis until at $\varepsilon=0$ (i.e. at ${\bf p}={\bf p}_0$) they merge and form the double imaginary eigenvalue $i\omega_0$ which further splits into two complex eigenvalues when $\varepsilon$ takes negative values. The opposite Krein signs  is a necessary and sufficient condition for the imaginary eigenvalues participating in the merging to leave the imaginary axis  \cite{BD2007,YS1975,Z2016}.  Below we demonstrate the Krein collision at the onset of the diffusionless AMRI by calculating the roots of the dispersion relation both analytically and numerically.

\subsection{Neutral stability curves}
Let $\delta:={\rm Ro}-{\rm Rb}{\rm S}^2$. In the Hamiltonian case ($\frac{1}{\rm Re}=0$, $\frac{1}{\rm Rm}=0$)  the dispersion relation  $p_0(\lambda):=\det({\bf H}_0-i^{-1}{\bf G}\lambda {\bf I})=0$
possesses a compact representation
\cite{FV95,OP1996,KSF2014b}
\be{idr}
p_0(\lambda)=4\delta^2+4\left(i\lambda-n+n{\rm S}^2\right)^2-\left(2\delta-(i\lambda-n)^2+n^2{\rm S}^2\right)^2=0.
\ee

If $\delta=0$, i.e. ${\rm Ro}={\rm Rb}{\rm S}^2$, then the equation \rf{idr} simplifies and its roots are \cite{KSF2014b}
\ba{r12}
&\lambda_{1,2}=-i(1+n)\pm i\sqrt{1-{\rm S}^2\left[1-(1+n)^2\right]},&\nn\\
&\lambda_{3,4}=-i(1-n)\pm i\sqrt{1-{\rm S}^2\left[1-(1-n)^2\right]}.&
\ea
The eigenvalues $\lambda_{1,2,3,4}$ are imaginary and simple for all $0< n \le 2$, if $0\le{\rm S}<1$. The equality ${\rm S}=1$ implies $\rm Ro=Rb$ and the existence of a double zero eigenvalue which is semi-simple at all $0\le n\le 2$ except $n=1$ where it has a Jordan block of order 2; the other two eigenvalue branches are formed by simple imaginary eigenvalues (marginal stability). At $\rm S>1$ complex eigenvalues originate (oscillatory instability), if
\be{bound}
{\rm S}>\frac{1}{\sqrt{1-(1-n)^2}}.
\ee
At the boundary of the domain \rf{bound} the eigenvalues are double imaginary with a Jordan block.

      \begin{figure}
    \begin{center}
    \includegraphics[angle=0, width=0.45\textwidth]{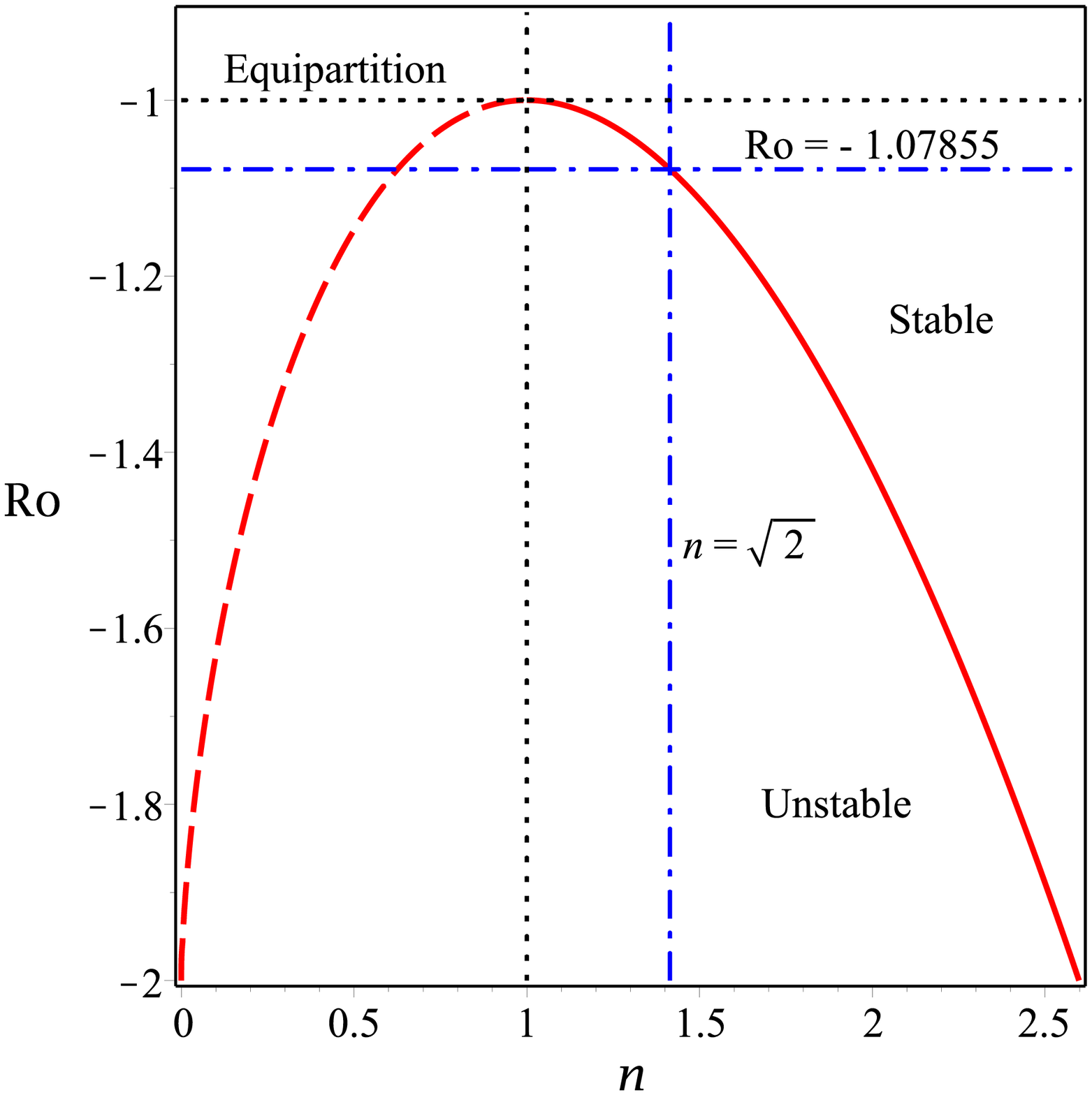}
    \includegraphics[angle=0, width=0.45\textwidth]{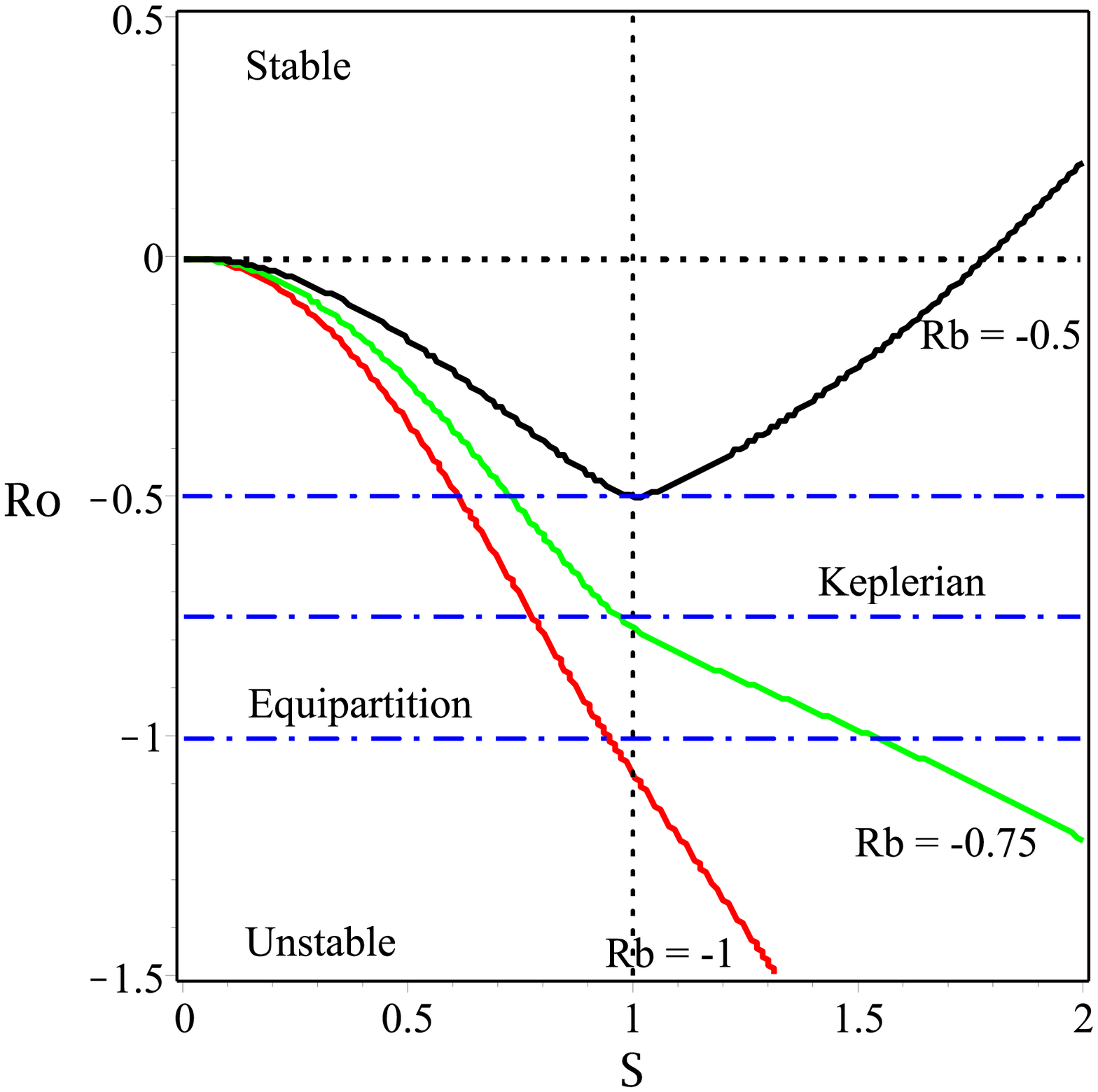}
    \end{center}
    \caption{(Left) Stability diagram in $(n,{\rm Ro})$-plane at $\rm S=1$ and $\rm Rb=-1$ according to the criterion \rf{unityd}. The dashed line shows the nonphysical branch of the neutral stability curve \rf{roc} corresponding to $0<n<1$. (Right) The critical value of $\rm Ro$ at the onset of the Hamilton-Hopf bifurcation as a function of $\rm S$ when $n=\sqrt{-2 \rm Rb}$ \cite{KSF2014b} at various values of $\rm Rb$.}
    \label{fig1}
    \end{figure}

In general, the instability corresponds to the negative discriminant of the polynomial \rf{idr}:
\ba{discrim}
&~&(n{\rm S})^6({\rm S}^2-1)^2 +2(n{\rm S})^4\left[({\rm S}^2+1)\delta^2+2({\rm S}^2-1)({\rm S}^2-2)\delta
+({\rm S}^2-1)^2(1-2{\rm S}^2)\right]\nn\\&+&(n{\rm S})^2\left[\delta^4+4(2{\rm S}^2+3)\delta^3-2(4{\rm S}^4+11({\rm S}^2-1))\delta^2+4({\rm S}^2-1)(5{\rm S}^2-3)\delta+({\rm S}^2-1)^2\right]\nn\\
&-&4\delta(\delta+1)^3({\rm S}^2-\delta-1)<0.
\ea

Following \cite{OP1996} we assume in \rf{discrim} that $n{\rm S}=c$, where $c=const$. Taking into account that $\delta={\rm Ro}-{\rm Rb}{\rm S}^2$ and then taking the limit ${\rm S}\rightarrow 0$, which obviously corresponds to the limit of $n \rightarrow \infty$, we find the following asymptotic expression for the instability condition \cite{OP1996}:
$$
(c^2+4{\rm Ro})(({\rm Ro}+1)^2+c^2)^2<0,
$$
or ${\rm S}^2<-\frac{4\rm Ro}{n^2}$, which yields \rf{opbh} at $\alpha=1$. At $n=0$ the inequality \rf{discrim} reduces to $\delta<-1$ which is exactly the diffusionless Michael criterion \rf{micr}.

Let us now assume that ${\rm S}=1$. Then, the inequality \rf{discrim} takes the form
\ba{unityd}
4n^4+(({\rm Ro}-{\rm Rb})^2+20({\rm Ro}-{\rm Rb})-8)n^2+4({\rm Ro}-{\rm Rb}+1)^3<0
\ea
and the dispersion relation at $\rm S=1$ factorizes as follows:
\be{rce}
\left. p_0(\lambda)\right|_{\rm S=1}=[\lambda^3+4in\lambda^2+4(1-n^2+{\rm Ro}-{\rm Rb})\lambda+8in({\rm Ro}-{\rm Rb})]\lambda=0.
\ee

The equality in  \rf{unityd} corresponds to the transition from marginal stability to oscillatory instability via the linear Hamilton-Hopf bifurcation, see Fig.~\ref{fig1}. At the marginal stability curve with $\rm S=1$ one of the eigenvalues $\lambda$ is always zero and simple,
another one is simple and imaginary and the last two form a double and imaginary eigenvalue with the Jordan block. At $\rm S=1$ and $\rm Rb=-1$ the critical value of the fluid Rossby number follows from \rf{unityd} and is equal to
\be{roc}
{\rm Ro}_c(n)=-2+\frac{\beta^{\frac{1}{3}}-n^2}{12}-\frac{n^2}{\beta^{\frac{1}{3}}}\left(18-\frac{n^2}{12}\right),
\ee
where
\be{rocb}
\beta(n)=-n^2\left(n^4+540n^2-5832-24\sqrt{3(n^2+27)^3}\right).
\ee

          \begin{figure}
    \begin{center}
    \includegraphics[angle=0, width=0.465\textwidth]{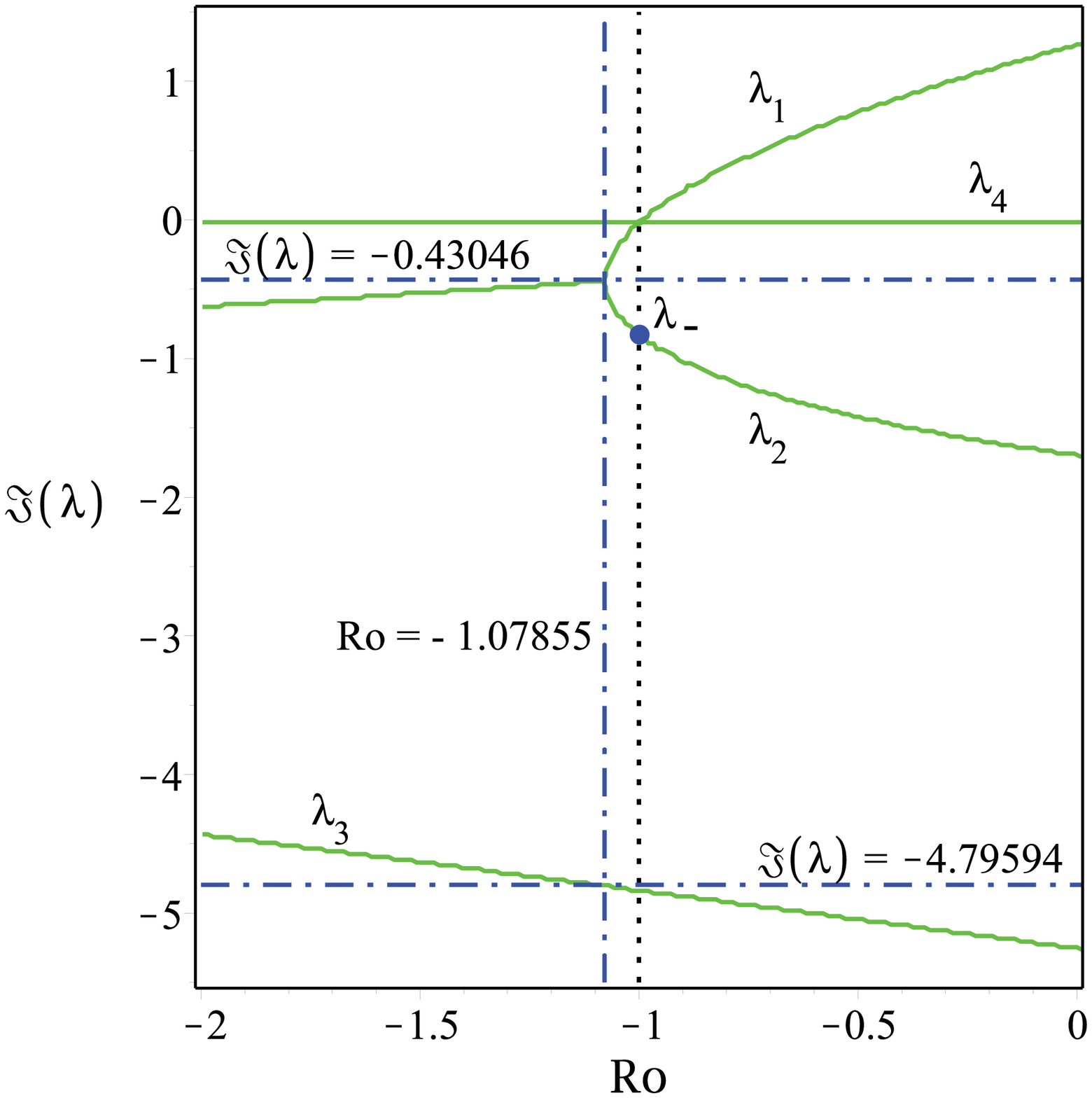}
    \includegraphics[angle=0, width=0.435\textwidth]{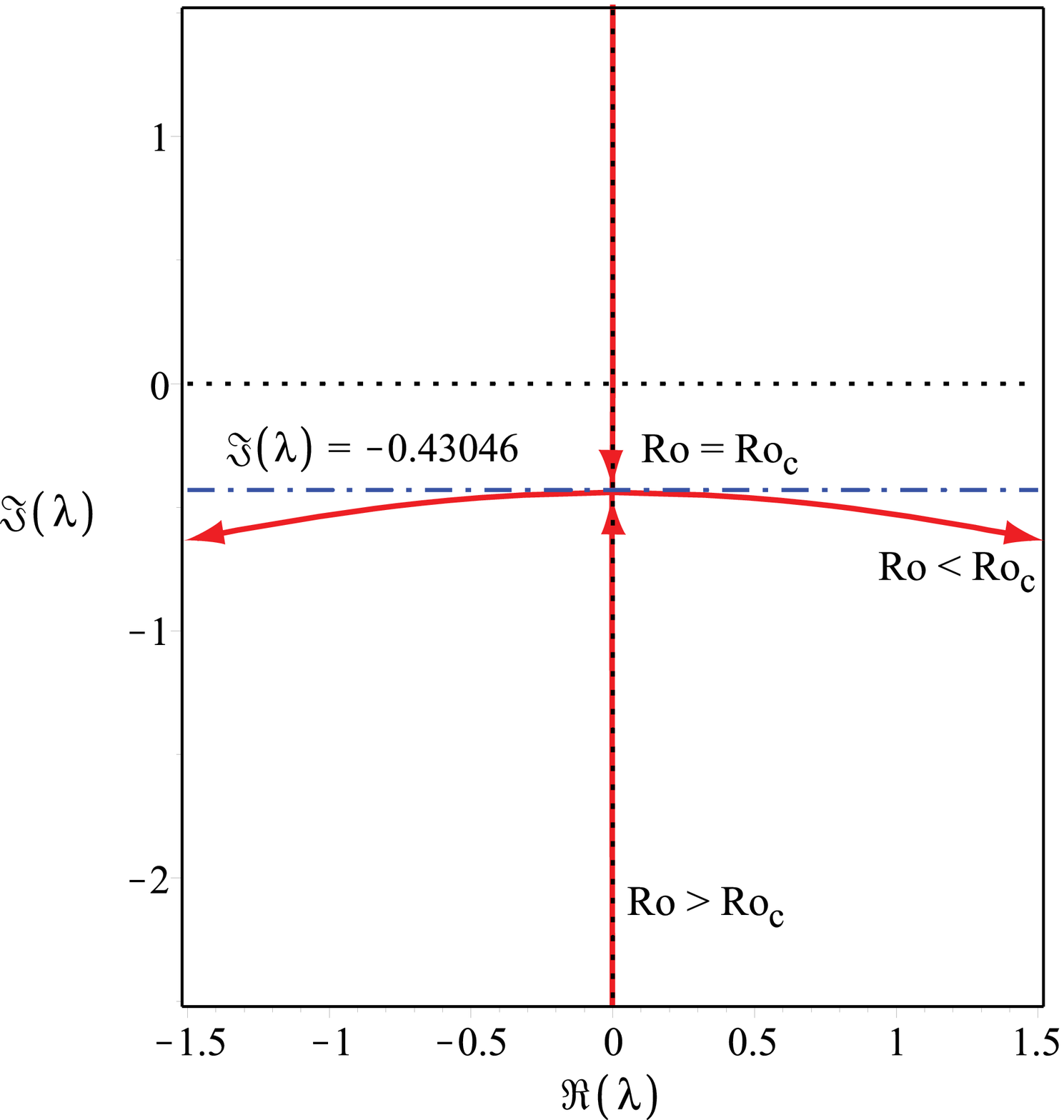}
    \end{center}
    \caption{(Left) Typical evolution of frequencies of the roots of the dispersion relation \rf{rce} as $\rm Ro$ is varied, shown for $\rm S=1$, ${\rm Rb}=-1$ and $n=\sqrt{2}$ that correspond to crossing the neutral stability curve along the vertical dash-dot line in Figure~\ref{fig1}(left). It demonstrates the Hamilton-Hopf bifurcation at ${\rm Ro}={\rm Ro}_c\approx-1.07855$ and the marginal stability of the Chandrasekhar energy equipartition solution at ${\rm Ro}=-1$. (Right) The same linear Hamilton-Hopf bifurcation shown in the complex plane: with the decrease in $\rm Ro$ two simple imaginary eigenvalues collide into a double imaginary eigenvalue with the Jordan block (an \textit{exceptional point} \cite{K2013dg}) that subsequently splits into two complex eigenvalues (oscillatory instability).}
    \label{fig2}
    \end{figure}

For example, at $n=\sqrt{2}$ Eq.~\rf{roc} yields
${\rm Ro}_c\approx-1.07855$, corresponding to the intersection of the two dash-dot lines in Fig.~\ref{fig1}(left). At this point of the curve \rf{roc} the eigenvalues are $\lambda_1=\lambda_2=\lambda_{c}$ (Fig.~\ref{fig2}), where
$$
\lambda_{c}=\frac{i\sqrt{2}}{34347}\left\{\frac{9\sqrt{87}+136}{4}\left[\beta\left(\sqrt{2}\right)\right]^{\frac{2}{3}}+
\frac{321\sqrt{87}-2782}{2}\left[\beta\left(\sqrt{2}\right)\right]^{\frac{1}{3}}-57245\right\}\approx-i0.43046,
$$
\be{flueig}
\lambda_{3}\approx-i4.79594,\quad \lambda_{4}=0.
\ee
Naturally, such explicit expressions for double imaginary eigenvalues can be obtained with the use of \rf{roc} and \rf{rocb} for any other value of $n$. The choice of $n$ does not influence the qualitative picture of eigenvalue interaction shown in Figure~\ref{fig2}. The value $n=\sqrt{2}$ is known to be optimal in several respects \cite{KSF2012,KSF2014b,SK2015}, which will be discussed further in the text.

\subsection{The Krein collision at the linear Hamilton-Hopf bifurcation threshold}
Although it is easy to evaluate the Krein sign of the imaginary eigenvalues shown in Figure~\ref{fig1} numerically,
it is instructive first to do it analytically in a particular case when $\rm Ro=Rb=-1$ and $\rm S=1$. Then, the eigenvalues are given explicitly by Eq.~\rf{r12}, which yields a double semi-simple zero eigenvalue $\lambda_0=0$ with the two linearly-independent eigenvectors ${\bf z}_1=(0,1,0,1)^T$ and ${\bf z}_2=(1,0,1,0)^T$ and the two imaginary eigenvalues $\lambda_{\pm}=-2i(n\pm1)$ with the eigenvectors ${\bf z}_+=\left(-i\alpha,\frac{-n}{2+n},\frac{in\alpha}{2+n},1\right)^T$ and ${\bf z}_-=\left(i\alpha,\frac{n}{2-n},\frac{in\alpha}{2-n},1\right)^T$, respectively, see Fig.~\ref{fig2}(left).

Notice that the eigenvalues $\lambda_+$ and $\lambda_-$ of Chandrasekhar's equipartition solution have the opposite Krein signs:
\be{kres}
\frac{[{\bf z}_+,{\bf z}_+]}{({\bf z}_+,{\bf z}_+)}=-\frac{2\alpha}{1+\alpha^2}\frac{2(n+1)^2}{1+(n+1)^2}<0, \quad \frac{[{\bf z}_-,{\bf z}_-]}{({\bf z}_-,{\bf z}_-)}=\frac{2\alpha}{1+\alpha^2} \frac{2(n-1)^2}{1+(n-1)^2}>0.
\ee

          \begin{figure}
    \begin{center}
    \includegraphics[angle=0, width=0.465\textwidth]{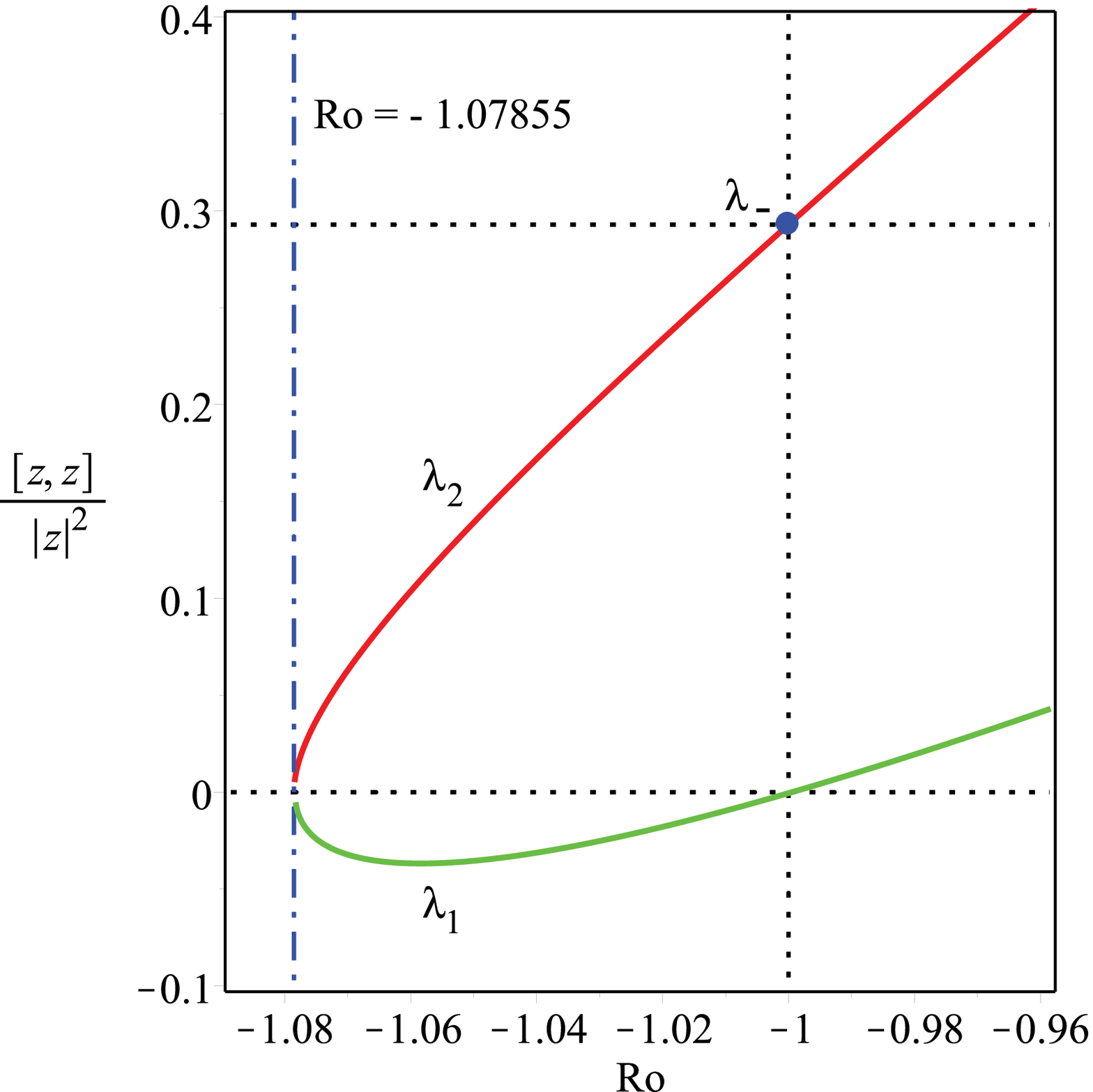}
    \includegraphics[angle=0, width=0.435\textwidth]{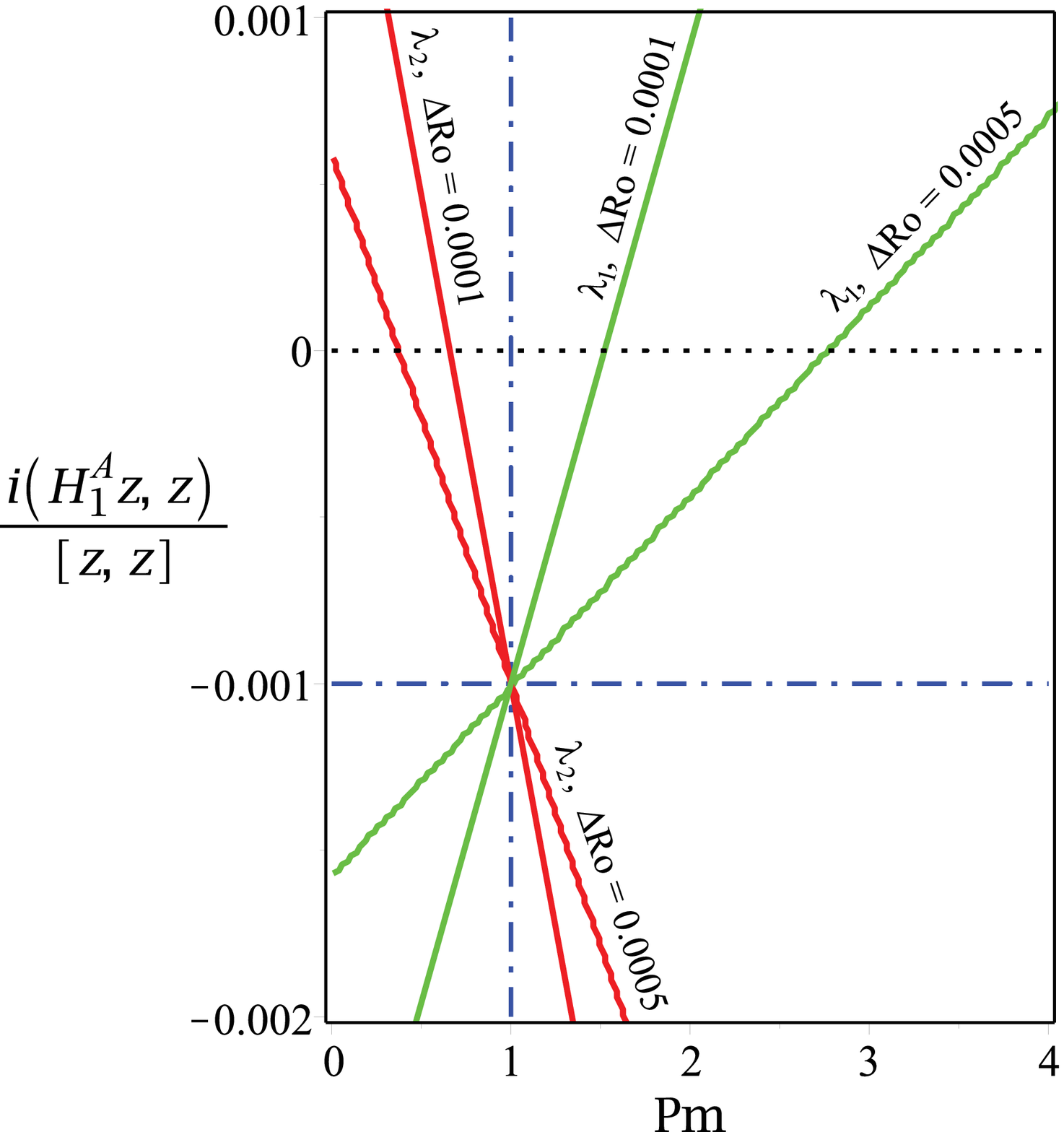}
    \end{center}
    \caption{For $\rm S=1$, ${\rm Rb}=-1$, $n=\sqrt{2}$, and $\alpha=1$ (left) the values of the normalized indefinite inner product $\frac{[{\bf z}, {\bf z}]}{({\bf z},{\bf z})}$ calculated with the eigenvectors at the imaginary eigenvalues $\lambda_1$ and $\lambda_2$ shown in Fig.~\ref{fig2}(left) that participate in the Hamilton-Hopf bifurcation at ${\rm Ro}={\rm Ro}_c\approx-1.07855$. For ${\rm Ro}_c<{\rm Ro}<-1$ the Krein sign of $\lambda_1$ is negative and the Krein sign of $\lambda_2$ is positive. (Right) For ${\rm Rm}=1000$ the values of the real increment $\delta\lambda^A$ to eigenvalues $\lambda_1$ with the negative Krein sign and to eigenvalues $\lambda_2$ with the positive Krein sign according to Eq.~\rf{dl}. The interval of negative increments (stability) around $\rm Pm=1$ becomes more narrow as $\Delta {\rm Ro}:={\rm Ro}-{\rm Ro}_c$ tends to zero. }
    \label{fig3}
    \end{figure}

For instance, at $n=\sqrt{2}$ we have $\frac{1+\alpha^2}{2\alpha}\frac{[{\bf z}_-,{\bf z}_-]}{({\bf z}_-,{\bf z}_-)}=1-\frac{\sqrt{2}}{2}\approx0.2929$, which implies that $\lambda_-$ has a positive Krein sign, see Fig.~\ref{fig3}(left).
The solid circle corresponding to $\lambda_-$ in Fig.~\ref{fig3}(left) belongs to the curve of the values of the normalized indefinite inner products $\frac{[{\bf z}, {\bf z}]}{({\bf z},{\bf z})}$ calculated on the eigenvectors at the eigenvalues of the branch marked as $\lambda_2$ in Fig.~\ref{fig2}(left). All imaginary eigenvalues $\lambda_2$ for ${\rm Ro}_c<{\rm Ro}<-1$ have positive Krein sign. In contrast, the eigenvalues of the branch $\lambda_1$ in Fig.~\ref{fig2}(left) have negative Krein sign on the same interval.

Therefore, the onset of the nonaxisymmetric oscillatory instability (or the diffusionless AMRI) is accompanied by the Krein collision of modes of positive and negative Krein sign, in accordance with the results of the section 3(a). The Krein sign is directly related to the sign of energy of a mode and the linear Hamilton-Hopf bifurcation is a collision of two imaginary eigenvalues of a Hamiltonian system with the opposite Krein (energy) signs \cite{K2013dg,YS1975,BD2007,IKS2009,Z2016}.

\section{Dissipation-induced instabilities of the double-diffusive system}
\subsection{Dissipative perturbation of simple imaginary eigenvalues}
The complex non-Hermitian matrix of the dissipative perturbation can be decomposed into its Hermitian and anti-Hermitian components: ${\bf H}_1={\bf H}_1^H+{\bf H}_1^A$, where
$${\bf H}_1^H=\frac{\rm S({\rm Pm}-1)}{2{\rm Rm}}\left(
                                                  \begin{array}{cccc}
                                                    0 & 0 & 0 & 1 \\
                                                    0 & 0 & -1 & 0 \\
                                                    0 & -1 & 0 & 0 \\
                                                    1 & 0 & 0 & 0 \\
                                                  \end{array}
                                                \right)$$ and
            $$
{\bf H}_1^A=\frac{1}{\rm Rm}\left(
              \begin{array}{cccc}
                0 & {\rm Pm} & 0 & -\frac{\rm S({\rm Pm}+1)}{2} \\
                -{\rm Pm}& 0 & \frac{\rm S({\rm Pm}+1)}{2} & 0 \\
                0 & -\frac{\rm S({\rm Pm}+1)}{2} & 4i\frac{{\rm Ro}-{\rm Rb}}{\alpha n } & 1 \\
                \frac{\rm S({\rm Pm}+1)}{2} & 0 & -1 & 0 \\
              \end{array}
            \right).
$$

At large $\rm Rm$ an increment $\delta\lambda$ to a simple imaginary eigenvalue $\lambda$ with an eigenvector $\bf z$ is given by a standard perturbation theory \cite{K2013dg,M1991,MO1995,BKMR94} as
\be{dl}
\delta\lambda=i\frac{\overline{\bf z}^T {\bf H}_1 {\bf z}}{\overline{\bf z}^T{\bf G}{\bf z}}=i\frac{({\bf H}_1{\bf z},{\bf z})}{[{\bf z},{\bf z}]}.
\ee

The increment $\delta\lambda^H=i\frac{({\bf H}_1^H{\bf z},{\bf z})}{[{\bf z},{\bf z}]}$ is obviously imaginary.
In particular, ${\bf H}_1^H=0$ at $\rm Pm=1$, i.e. the frequencies are not affected by the Hermitian component of the dissipative perturbation if the contributions from viscosity and resistivity are equal.

In contrast, the increment $\delta\lambda^A=i\frac{({\bf H}_1^A{\bf z},{\bf z})}{[{\bf z},{\bf z}]}$ is real. For instance,
the eigenvalues $\lambda_+$ and $\lambda_-$ of Chandrasekhar's equipartition solution acquire the following increments:
\be{dla}
\delta \lambda_{\pm}^A=-\frac{\rm Pm+1}{\rm 2Rm}=-\frac{1}{h}:=-\frac{1}{2}\left(\frac{1}{\rm Re}+\frac{1}{\rm Rm}\right),\quad \delta \lambda_{\pm}^H=0,
\ee
where $h$ is the harmonic mean of the two Reynolds numbers.

\subsection{Weak ohmic diffusion destabilizes positive energy waves at low $\rm Pm$}
In the close vicinity of the critical Rossby number of the Hamilton-Hopf bifurcation $\rm Ro_c \approx -1.07855$ the real increment $\delta\lambda^A$ to imaginary eigenvalues $\lambda_1$ with negative Krein sign and  $\lambda_2$ with positive Krein sign are shown in Fig.~\ref{fig3}(right) for the fixed $\rm Rm=10^3$ and varying $\rm Pm$ (the fluid Reynolds number is calculated as $\rm Re=Rm/Pm$).

The eigenvalues with the \textit{negative} Krein sign become dissipatively-destabilized when ${\rm Pm}>1$, i.e. when the losses due to viscosity of the fluid exceed the ohmic losses, cf. \cite{A1973}. Remarkably, the eigenvalues with the \textit{positive} Krein sign can also acquire positive growth rates. However, this happens at $\rm Pm < 1$ when the electrical resistivity prevails over the kinematic viscosity. Indeed, the destabilizing influence of the kinematic viscosity of the fluid on negative energy waves is well-known in hydrodynamics \cite{A1973,Yih1961,TSL2013,BD2007}, which therefore places the dissipation-induced instability at $\rm Pm >1$ and $|\rm Ro-{\rm Ro}_c|\ll1$ into an established context. The destabilization of positive energy modes was noticed in the context of solid mechanics, in particular, in  gyroscopic systems with damping and non-conservative positional (or circulatory, or curl \cite{BS2012}) forces in \cite{KV10,K2009,K2013pt,K2013dg}. Radiative dissipation due to emission of electromagnetic, acoustic, and gravitational waves is a well-known reason for instability of modes of positive energy in hydrodynamics and plasma physics \cite{L1996,ORT1986,C1984,LD1977}. To the best of our knowledge, the dissipative destabilization of the positive energy modes due to ohmic losses has not been previously reported in MHD.

The interval of negative real increments in Fig.~\ref{fig3}(right) decreases with decreasing deviation from the critical value of the Rossby number at the Hamilton-Hopf bifurcation, i.e. as $\Delta{\rm Ro}={\rm Ro}-{\rm Ro}_c$ tends to zero. When $\Delta{\rm Ro}=0$, the stable interval reduces to the single value: $\rm Pm=1$. Hence, weak ohmic diffusion (weak kinematic viscosity) destabilizes positive (negative) energy waves at $\rm Pm<1$ ($\rm Pm>1$), if $|\rm Ro-{\rm Ro}_c|$ is sufficiently small.

          \begin{figure}
    \begin{center}
    \includegraphics[angle=0, width=0.45\textwidth]{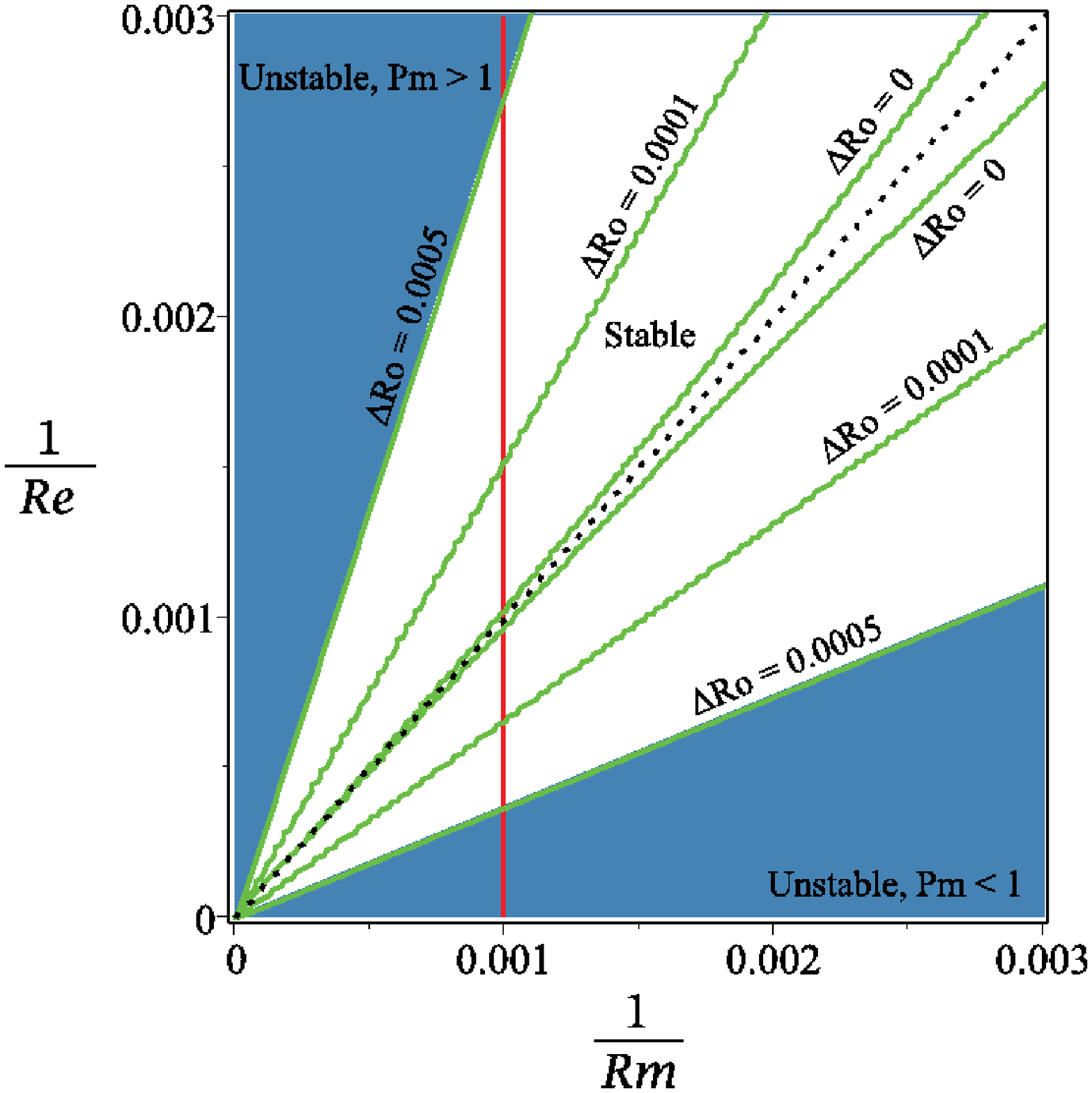}
    \includegraphics[angle=0, width=0.45\textwidth]{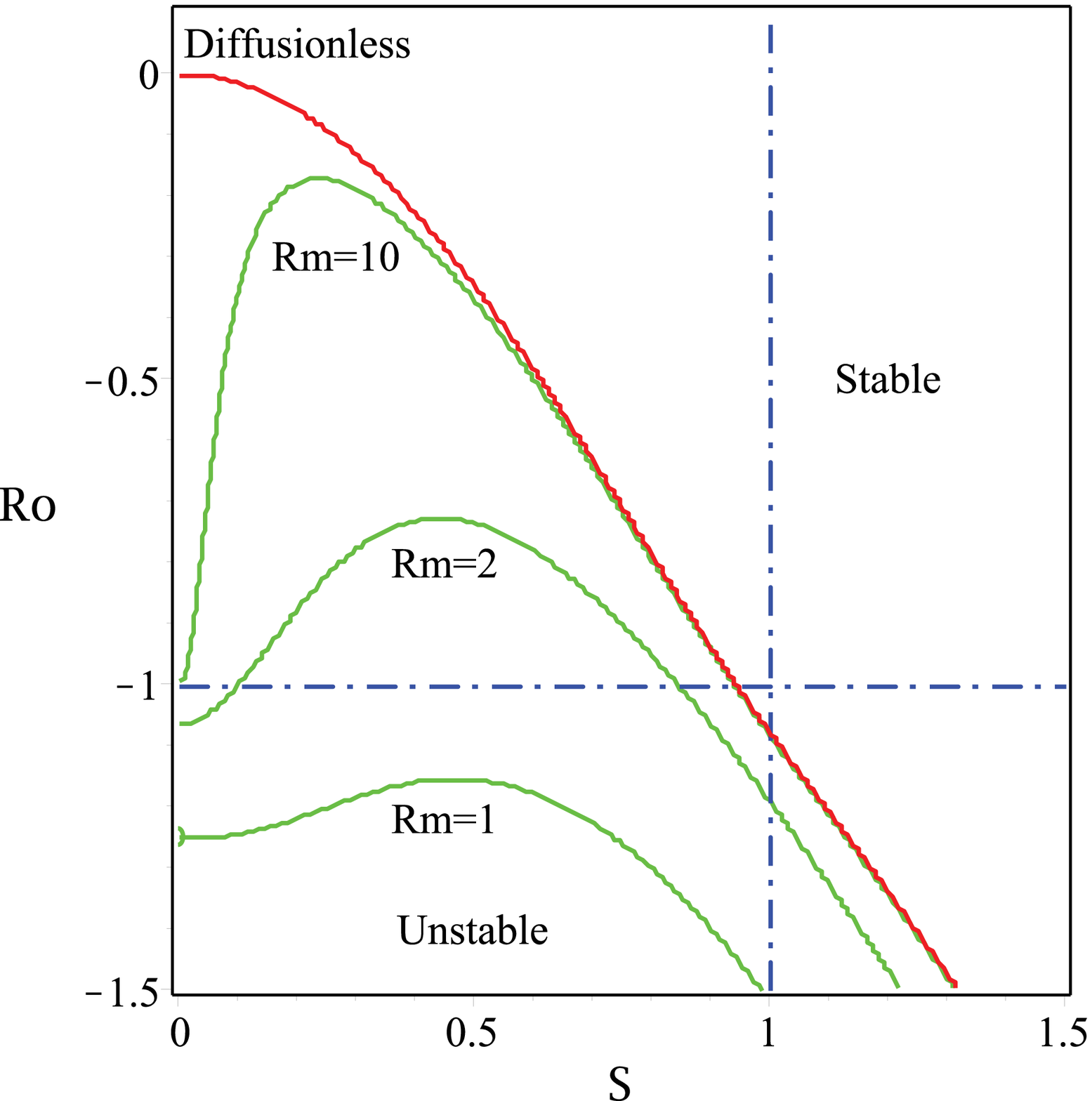}
    \end{center}
    \caption{(Left) For $\rm S=1$, ${\rm Rb}=-1$, and $n=\sqrt{2}$ the neutral stability curves in the plane $({\rm Rm}^{-1},{\rm Re}^{-1})$ of the inverse magnetic and fluid Reynolds numbers corresponding to different values of $\Delta {\rm Ro}:={\rm Ro}-{\rm Ro}_c$. The stability domain has a shape of an angular sector at $\Delta {\rm Ro}>0$ and a cusp at $\Delta {\rm Ro}=0$ with the single tangent line $\rm Pm=1$,  cf. Fig~\ref{fig3}(right). (Right) The neutral stability curves for ${\rm Rb}=-1$, $n=\sqrt{2}$, and $\rm Re=Rm$ in the $({\rm S},{\rm Ro})$-plane at various values of $\rm Rm$. }
    \label{fig4}
    \end{figure}

\subsection{Diffusionless and double-diffusive criteria are connected at $\rm Pm=1$}

We complement the sensitivity analysis of eigenvalues of the diffusionless Hamiltonian eigenvalue problem with respect to a double-diffusive perturbation with the direct computation of the stability boundaries based on the algebraic Bilharz stability criterion. The Bilharz criterion \cite{B1944} guarantees localization of all the roots of a complex polynomial of degree $n$ to the left of the imaginary axis in the complex plane provided that all principal minors of even order of the $2n\times2n$ Bilharz matrix composed of the real and imaginary parts of the coefficients of the polynomial are positive \cite{K2013dg}.

Applying the Bilharz criterion to the characteristic polynomial of the eigenvalue problem \rf{fep} we plot the neutral stability curves in the plane of the inverse Reynolds numbers ${\rm Rm}^{-1}$ and ${\rm Re}^{-1}$ at various values of $\Delta{\rm Ro}={\rm Ro}-{\rm Ro}_c$, where ${\rm Ro}_c$ is defined in \rf{roc}, when $\rm S=1$, $\rm Rb=-1$, and $n=\sqrt{2}$, Fig.~\ref{fig4}(left). Note that the diagonal ray corresponding to $\rm Pm=1$ always stays in the stability domain when $\Delta{\rm Ro}\ge 0$ and is the only tangent line to the stability boundary at the cuspidal point at the origin when ${\rm Ro}={\rm Ro}_c$. Moreover, at ${\rm Ro}={\rm Ro}_c$ and $\rm Re=Rm$ the spectrum of the double-diffusive system with $\rm S=1$ and $\rm Rb=-1$ contains the double complex eigenvalues (exceptional points \cite{K2013dg})
\be{depm}
\lambda_d=\lambda_c(n)-\frac{1}{\rm Rm}.
\ee
The imaginary eigenvalue $\lambda_c(n)$ is given in \rf{flueig} for the particular case of $n=\sqrt{2}$ .

Approaching the origin along the ray $\rm Pm=1$ means letting the Reynolds numbers tend to infinity with their ratio being kept equal to unity. Fig.~\ref{fig4}(right) demonstrates that in the limit ${\rm Re}={\rm Rm}\rightarrow \infty$ the neutral stability curve of the double-diffusive system approaches the threshold of instability of the diffusionless system from below. The instability domain of the double-diffusive system always remains smaller than in the diffusionless case. As a consequence, the Chandrasekhar equipartition solution $({\rm Ro}={\rm Rb}=-1, {\rm S}=1)$, being stable in the diffusionless case, remains stable at $\rm Pm=1$ no matter what the value of the Reynolds numbers is, Fig.~\ref{fig4}(right).

          \begin{figure}
    \begin{center}
    \includegraphics[angle=0, width=0.43\textwidth]{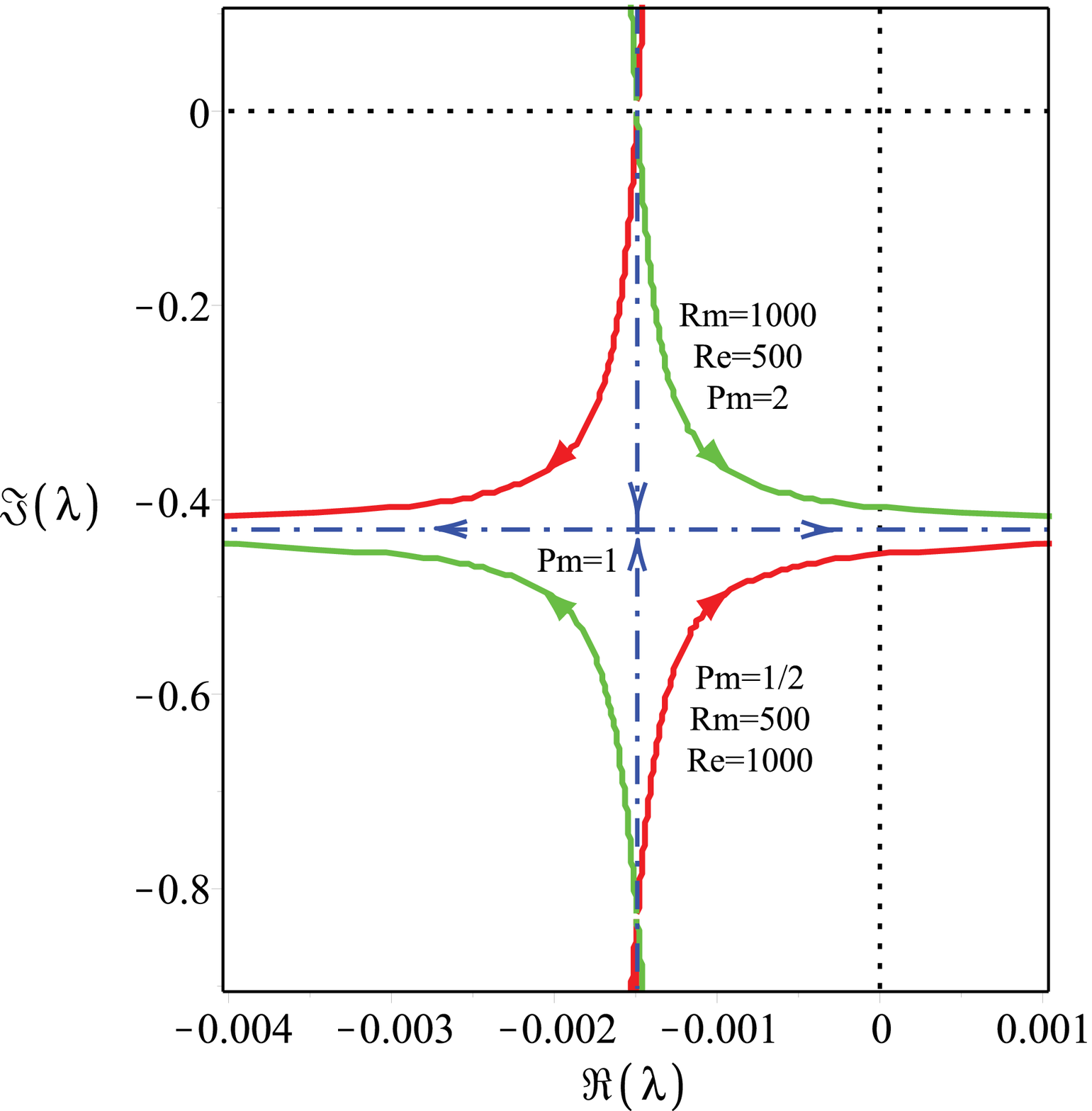}
    \includegraphics[angle=0, width=0.43\textwidth]{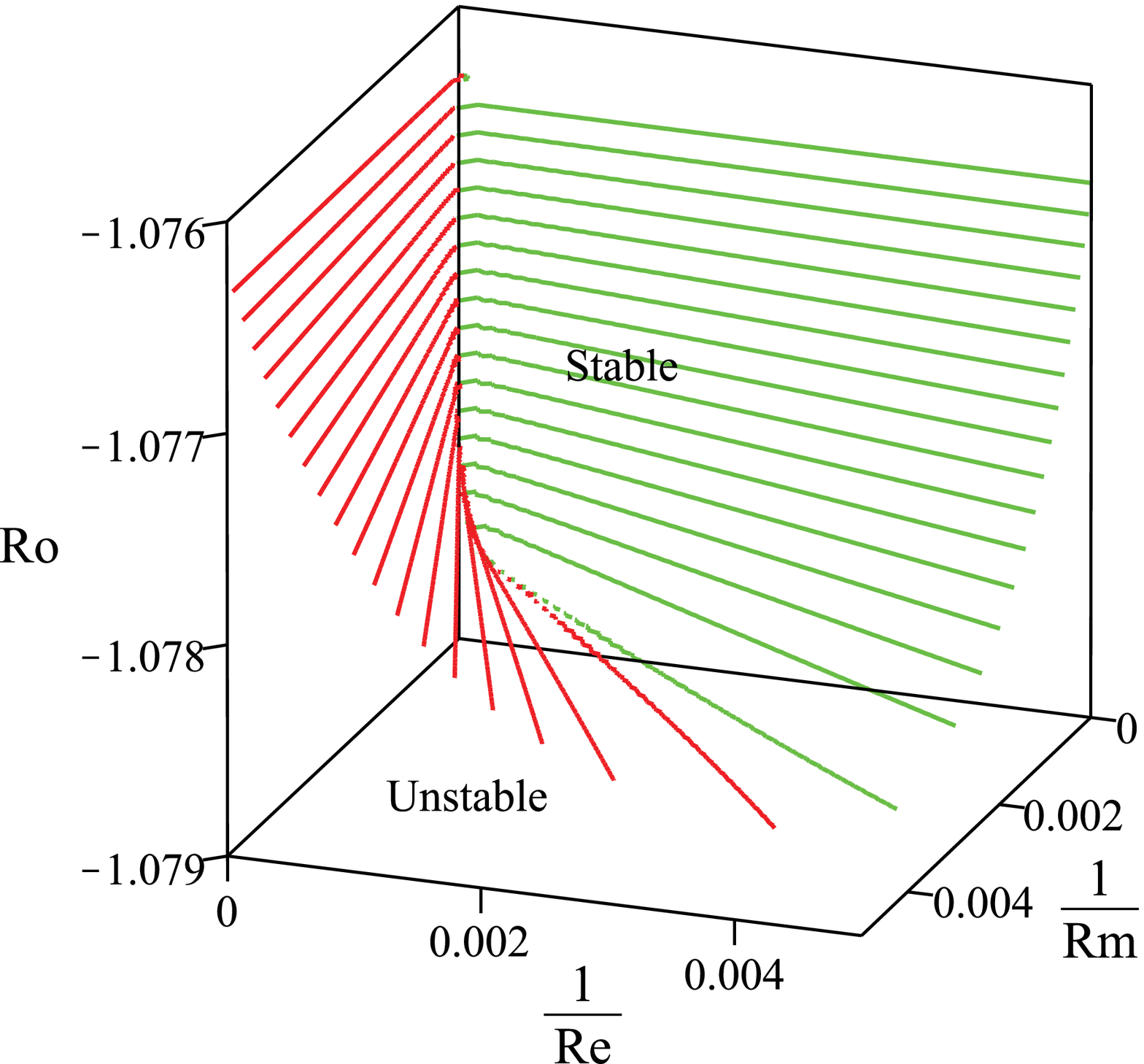}
    \end{center}
    \caption{(Left) At ${\rm Rb}=-1$, $\rm S=1$, and $n=\sqrt{2}$  the dash-dot lines show interaction of complex eigenvalues with negative real parts in the complex $\lambda$-plane with the decrease in $\rm Ro$ when ${\rm Re}={\rm Rm}=h=\frac{2}{\frac{1}{500}+\frac{1}{1000}}$, i.e. $\rm Pm=1$. At ${\rm Ro}={\rm Ro}_c$ the eigenvalues merge into the double complex eigenvalue \rf{depm}. The quasi-hyperbolic curves demonstrate the imperfect merging of modes (the avoided crossing) such that the mode with positive Krein  (energy) sign becomes unstable at $\rm Pm<1$ and the mode with negative Krein (energy) sign is unstable at $\rm Pm>1$. (Right) The neutral stability surface represented by the contours ${\rm Ro}=const.$ in the $({\rm Re}^{-1},{\rm Rm}^{-1},{\rm Ro})$-space has a ``Whitney umbrella'' singular point at $(0,0,{\rm Ro}_c)$ yielding a cusp in the cross-section ${\rm Ro}={\rm Ro}_c$ with the single tangent line   $\rm Pm=1$. }
    \label{fig5}
    \end{figure}

Indeed, in the case when ${\rm Ro}={\rm Rb}{\rm S}^2$ and ${\rm Re}={\rm Rm}$, the roots of the characteristic polynomial of the eigenvalue problem \rf{fep} can be found explicitly
\ba{rm12}
&\lambda_{1,2}=-i(n+1)-\frac{1}{\rm Rm}\pm i\sqrt{1-{\rm S}^2[1-(n+1)^2]},&\nn\\
&\lambda_{3,4}=-i(n-1)-\frac{1}{\rm Rm}\pm i\sqrt{1-{\rm S}^2[1-(n-1)^2]}.&
\ea

The eigenvalues \rf{rm12} are just the eigenvalues \rf{r12} that are shifted by dissipation to the left in the complex plane (asymptotic stability). This fact agrees perfectly with the result of Bogoyavlenskij \cite{Bog2004} who found at ${\rm Pm}=1$ exact unsteady energy equipartition solutions of the viscous and resistive incompressible MHD equations that relax with the growth rate equal to $-\frac{1}{\rm Re}=-\frac{1}{\rm Rm}<0$ to the ideal and steady Chandrasekhar equipartition equilibria \cite{Chandra56}. Notice also that even earlier Lerner and Knobloch  reported a `cooperative, accelerated decay' of solutions at ${\rm Pm}=1$ in the study of stability of the magnetized plane Couette flow \cite{LK1985}.

Well-known is a similar result on the secular instability of the Maclaurin spheroids due to both fluid viscosity and gravitational radiation reaction\footnote{The dissipation-induced instability of Maclaurin spheroids due to emission of gravitational waves is known as the Chandrasekhar-Friedman-Schutz (CFS) instability \cite{L1996}.} when the value of the critical eccentricity of the meridional section at the onset of instability in the ideal case is attained only when the ratio of the two dissipation mechanisms is exactly 1 \cite{LD1977,L1996}.

\subsection{Double-diffusive instability at $\rm Pm\ne1$ and arbitrary $\rm Re$ and $\rm Rm$  }

\subsubsection{Unfolding the Hamilton-Hopf bifurcation in the vicinity of ${\rm Pm}=1$}

Along $\rm Re=Rm>0$ the variation of $\rm Ro$ at fixed $\rm Rb=-1$, $\rm S=1$, and $n$ is accompanied by a bifurcation at ${\rm Ro}={\rm Ro}_c$ of the double complex eigenvalue \rf{depm} with negative real part equal to $-{\rm Rm}^{-1}$, Fig.~\ref{fig5}(left). Effectively, at $\rm Pm=1$ dissipation shifts the Hamilton-Hopf bifurcation to the left in the complex plane. For this reason, the oscillatory instability in the double-diffusive system with equal viscosity and resistivity occurs through the classical Hopf bifurcation at ${\rm Ro}(\rm Rm)<{\rm Ro}_c$ with ${\rm Ro}(\rm Rm)$ tending to ${\rm Ro}_c$ as $\rm Rm \rightarrow \infty$.

In the case when the magnetic Prandtl number slightly deviates from the value $\rm Pm=1$, the
shifted Hamilton-Hopf bifurcation unfolds into a couple of quasi-hyperbolic eigenvalue branches passing close to each other in an avoided crossing centered at an exceptional point $\lambda_d$ of the family \rf{depm} with real part equal to $-h^{-1}$, where $h=\frac{2}{\frac{1}{\rm Re}+\frac{1}{\rm Rm}}$ is the harmonic mean of the  fluid and magnetic Reynolds numbers, $\rm Re\ne Rm$, Fig.~\ref{fig5}(left).

                \begin{figure}
    \begin{center}
    \includegraphics[angle=0, width=0.45\textwidth]{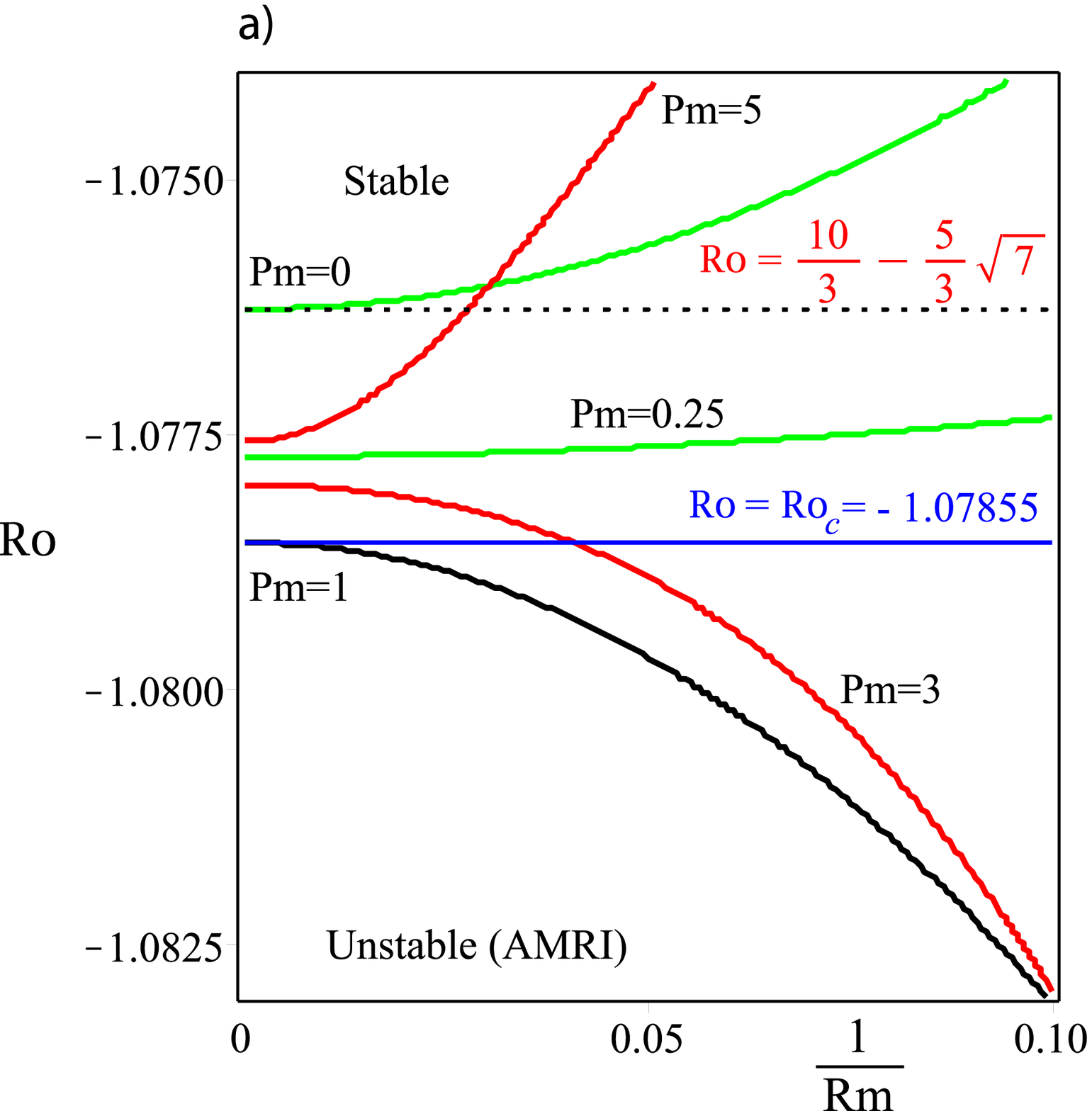}
    \includegraphics[angle=0, width=0.45\textwidth]{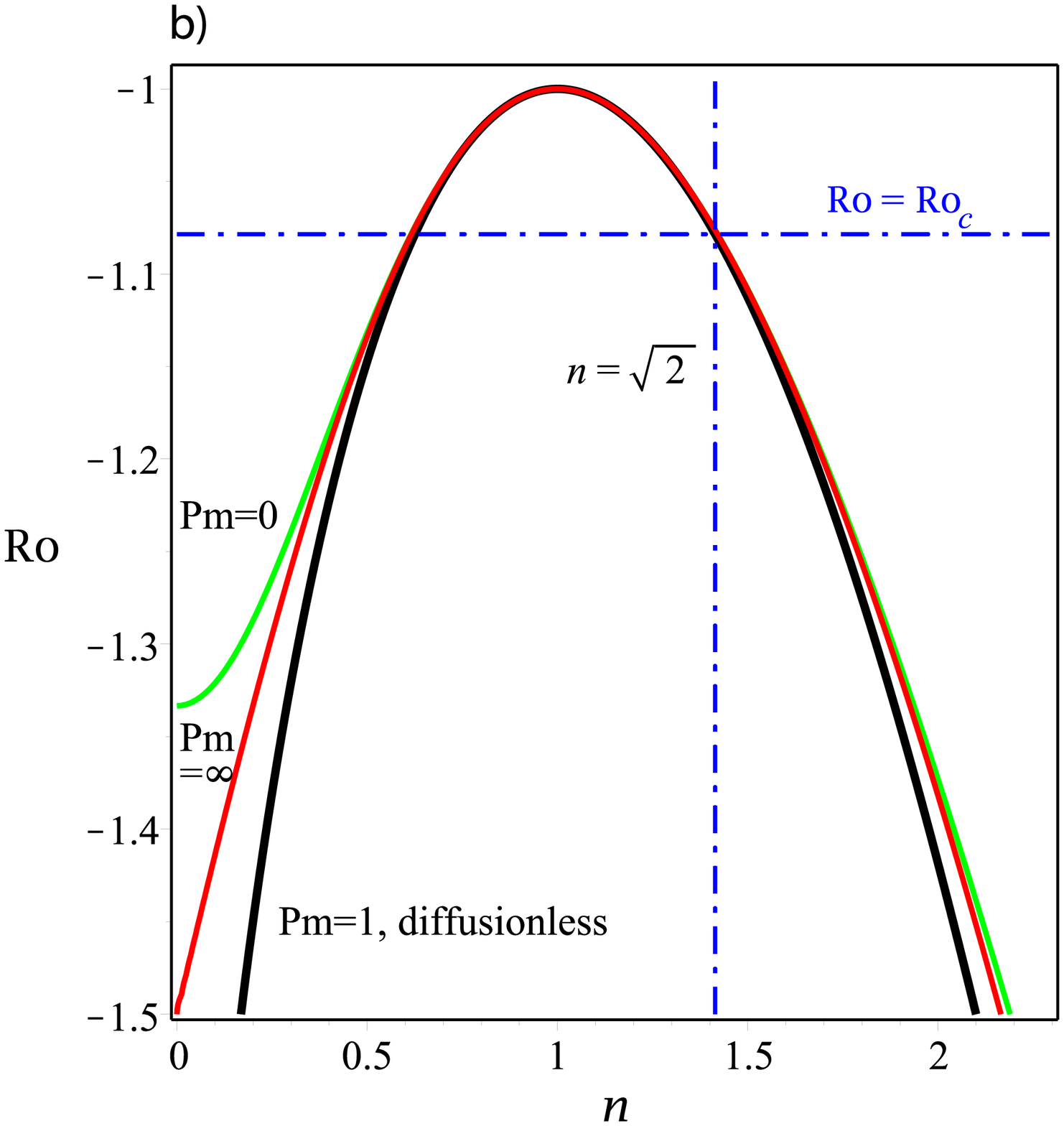}
    \includegraphics[angle=0, width=0.45\textwidth]{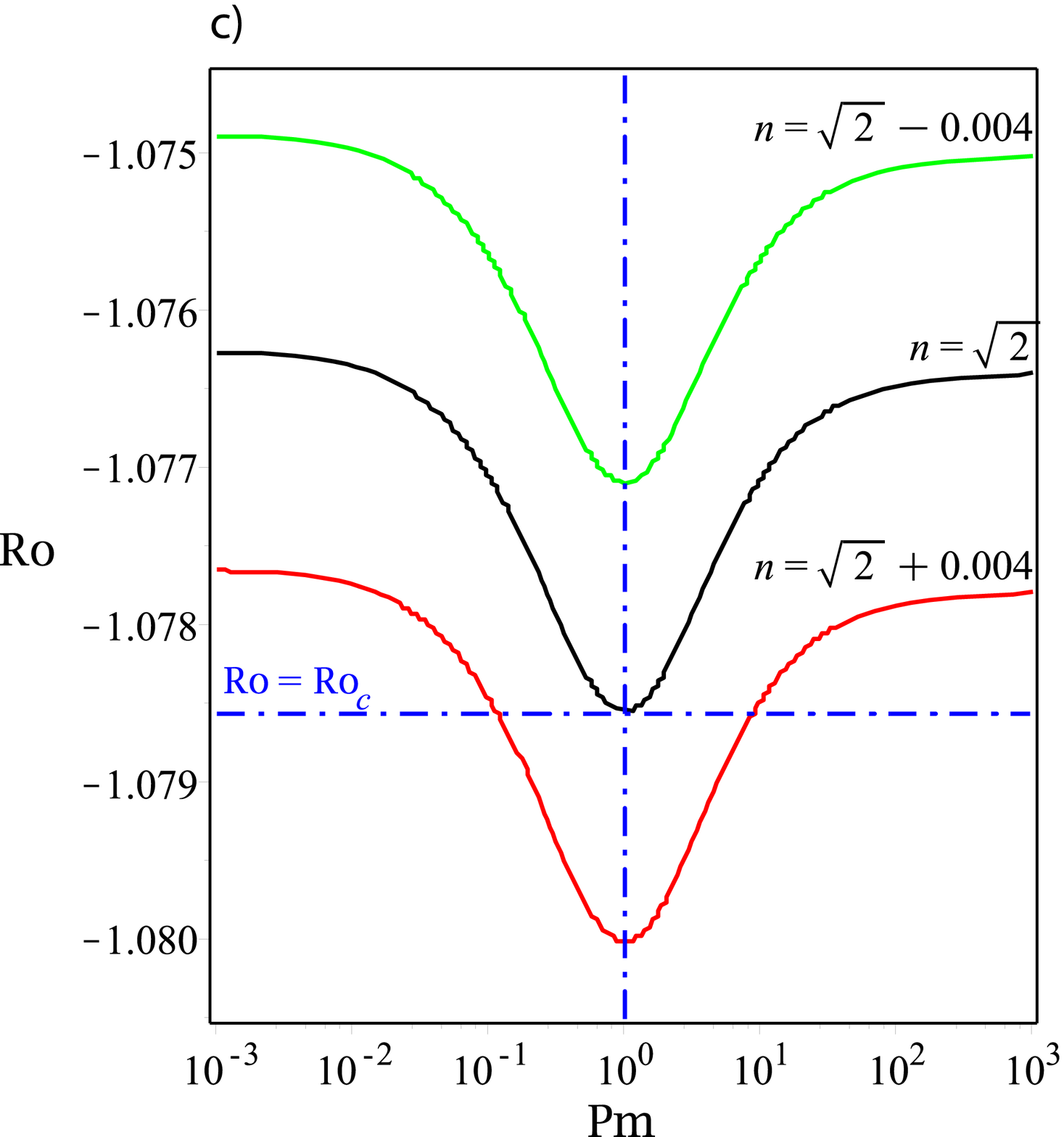}
    \includegraphics[angle=0, width=0.45\textwidth]{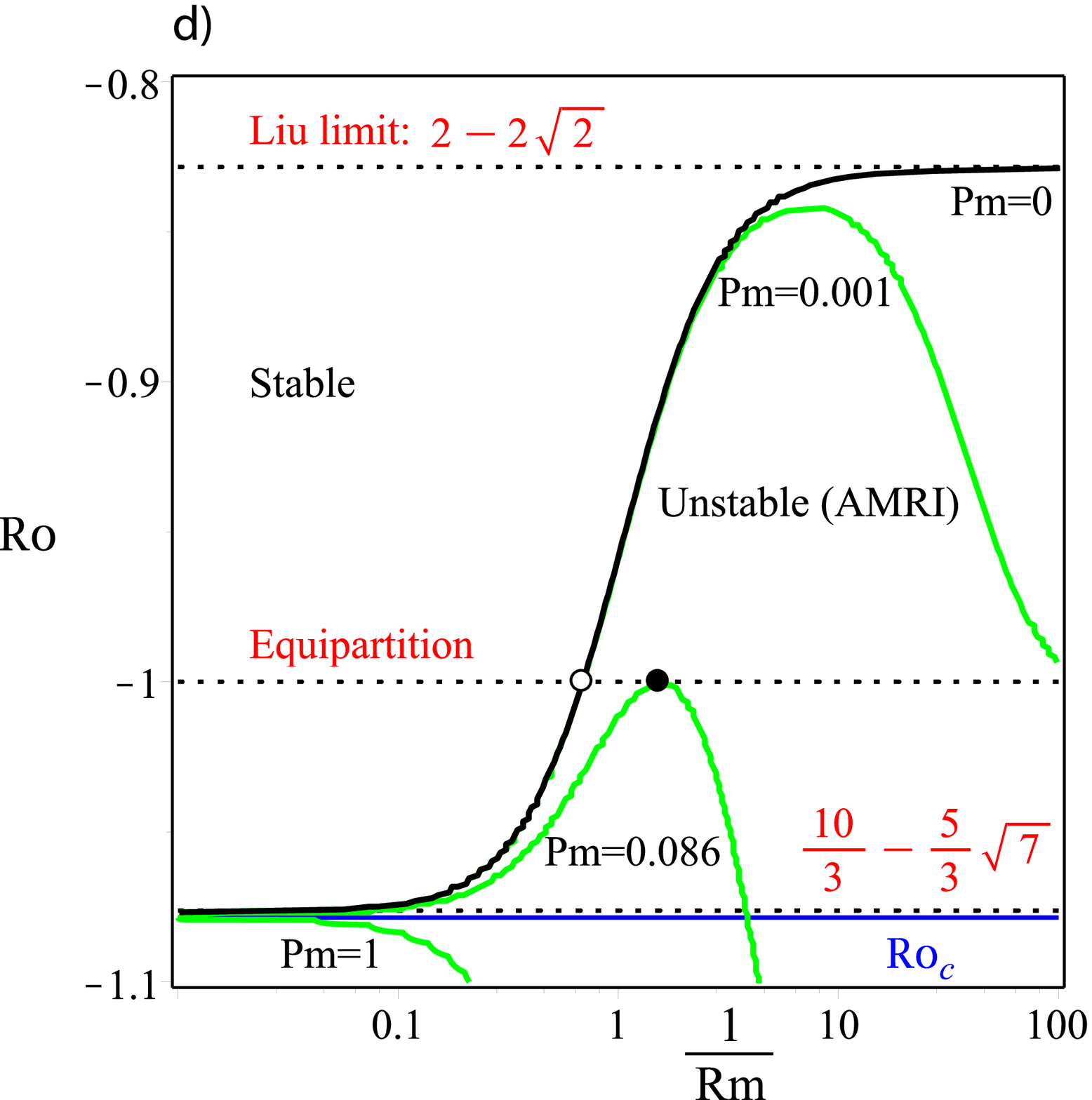}
    \end{center}
    \caption{(a) For ${\rm Rb}=-1$, $\rm S=1$, $n=\sqrt{2}$, and ${\rm Re}={\rm Rm}/{\rm Pm}$ the neutral stability curves in the $({\rm Rm}^{-1},{\rm Ro})$-plane demonstrating that the limit of the critical value of ${\rm Ro}$ as ${\rm Rm} \rightarrow \infty$ depends on $\rm Pm$ and attains its minimum ${\rm Ro}_c$ at $\rm Pm=1$. (b) The limit of the critical value of ${\rm Ro}$ at ${\rm Rb}=-1$, $\rm S=1$, and ${\rm Re}={\rm Rm}/{\rm Pm}$ as  ${\rm Rm} \rightarrow \infty$ plotted as a function of $n$ for (black) $\rm Pm=1$, (green) $\rm Pm=0$, and (red) $\rm Pm \rightarrow \infty$. The limit coincides with the stability boundary of the dissipationless case only at $\rm Pm=1$, independently on the choice of $n$. Similarly, at any $\rm Pm \ne 1$ the finite discrepancy between the dissipationless stability curve and the neutral stability curve in the limit of vanishing dissipation exists for all physically relevant values $n>1$. (c) The limit of the critical ${\rm Ro}$ given by Eq.~\rf{limpm} always has a minimum at $\rm Pm=1$. (d) For ${\rm Rb}=-1$, $\rm S=1$, $n=\sqrt{2}$, and ${\rm Re}={\rm Rm}/{\rm Pm}$ the neutral stability curves at various ${\rm Pm}\in [0,1]$ demonstrating that the maximal critical values of $\rm Ro$ do not exceed the Liu limit $2-2\sqrt{2}$ that is attained only at $\rm Pm=0$ in the limit of $\rm Rm \rightarrow 0$.}
    \label{fig6}
    \end{figure}

The unfolding of the eigenvalue crossing into the avoided crossing can happen in two different ways depending on the sign of $\rm Pm-1$. At $\rm Pm < 1$ ($\rm Pm > 1$) the complex eigenvalues stemming from the imaginary eigenvalues of the diffusionless system with positive (negative) Krein sign form a branch that bends to the right and crosses the imaginary axis at some ${\rm Ro}(\rm Re, Rm) \ne {\rm Ro}_c$, Fig.~\ref{fig5}(left), cf. \cite{C1984}. The critical values ${\rm Ro}(\rm Re, Rm)$ of the double-diffusive system live on the surface in the $({\rm Re}^{-1},{\rm Rm}^{-1},{\rm Ro})$-space that has a self-intersection along the $\rm Ro$-axis, Fig.~\ref{fig5}(right). The angle of the self-intersection tends to zero as ${\rm Ro} \rightarrow {\rm Ro}_c$ and at the point $(0,0,{\rm Ro}_c)$ the surface has a singularity known as the Whitney umbrella\footnote{The normal form of a surface in the $Oxyz$-space that has the Whitney umbrella singular point at the origin is given by the equation $zy^2=x^2$ \cite{A1971,L2003,B1956}.} \cite{L2003,KV10,A1971}.

In the vicinity of the $\rm Ro$-axis the instability threshold is effectively a ruled surface \cite{B1956}, where the slope of each ruler is determined by $\rm Pm$. Letting the Reynolds numbers tend to infinity while keeping the magnetic Prandtl number fixed means that the $\rm Ro$-axis is approached
in the $({\rm Re}^{-1},{\rm Rm}^{-1},{\rm Ro})$-space along a ruler corresponding to this value of $\rm Pm$. Generically, for all values of $\rm Pm$ except $\rm Pm=1$ a ruler leads to a limiting value of ${\rm Ro}$ that exceeds ${\rm Ro}_c$ and thus extends the instability interval of the fluid Rossby numbers with respect to that of the diffusionless system, as is visible in Fig.~\ref{fig5}(right) and Fig.~\ref{fig6}(a). The plane $\rm Pm=1$ divides the neutral stability surface in the vicinity of ${\rm Ro}={\rm Ro}_c$ into two parts corresponding to positive energy modes destabilized by the dominating ohmic diffusion at $\rm Pm<1$ and to negative energy modes destabilized by the dominating fluid viscosity at $\rm Pm >1$, Fig.~\ref{fig5}(right). The ray determined by the conditions $\rm Re=Rm>0$, ${\rm Ro}={\rm Ro}_c$ belongs to the stability domain of the double-diffusive system and contains exceptional points \rf{depm} that determine\footnote{This was anticipated by Jones \cite{Jones88}: "It is quite common for an eigenvalue
which is moving steadily towards a positive growth rate to suffer a sudden change of direction and subsequently
fail to become unstable; similarly, it happens that modes which initially become more stable
as [the Reynolds number] increases change direction and subsequently achieve instability. It is believed
that these changes of direction are due to the nearby presence of multiple-eigenvalue points."} behaviour of eigenvalues shown in Fig.~\ref{fig5}(left).

            \begin{figure}
    \begin{center}
    \includegraphics[angle=0, width=0.45\textwidth]{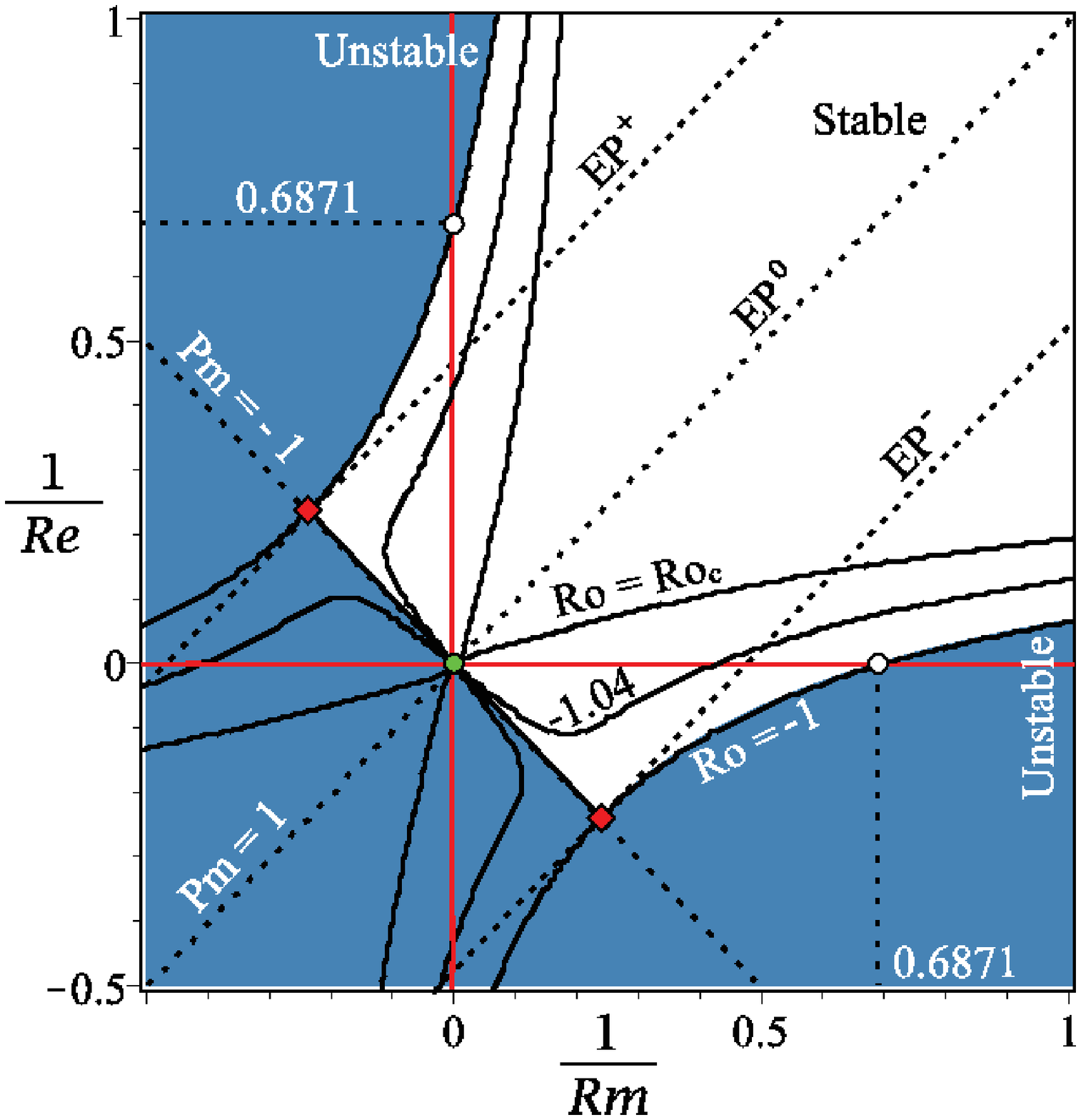}
    \includegraphics[angle=0, width=0.45\textwidth]{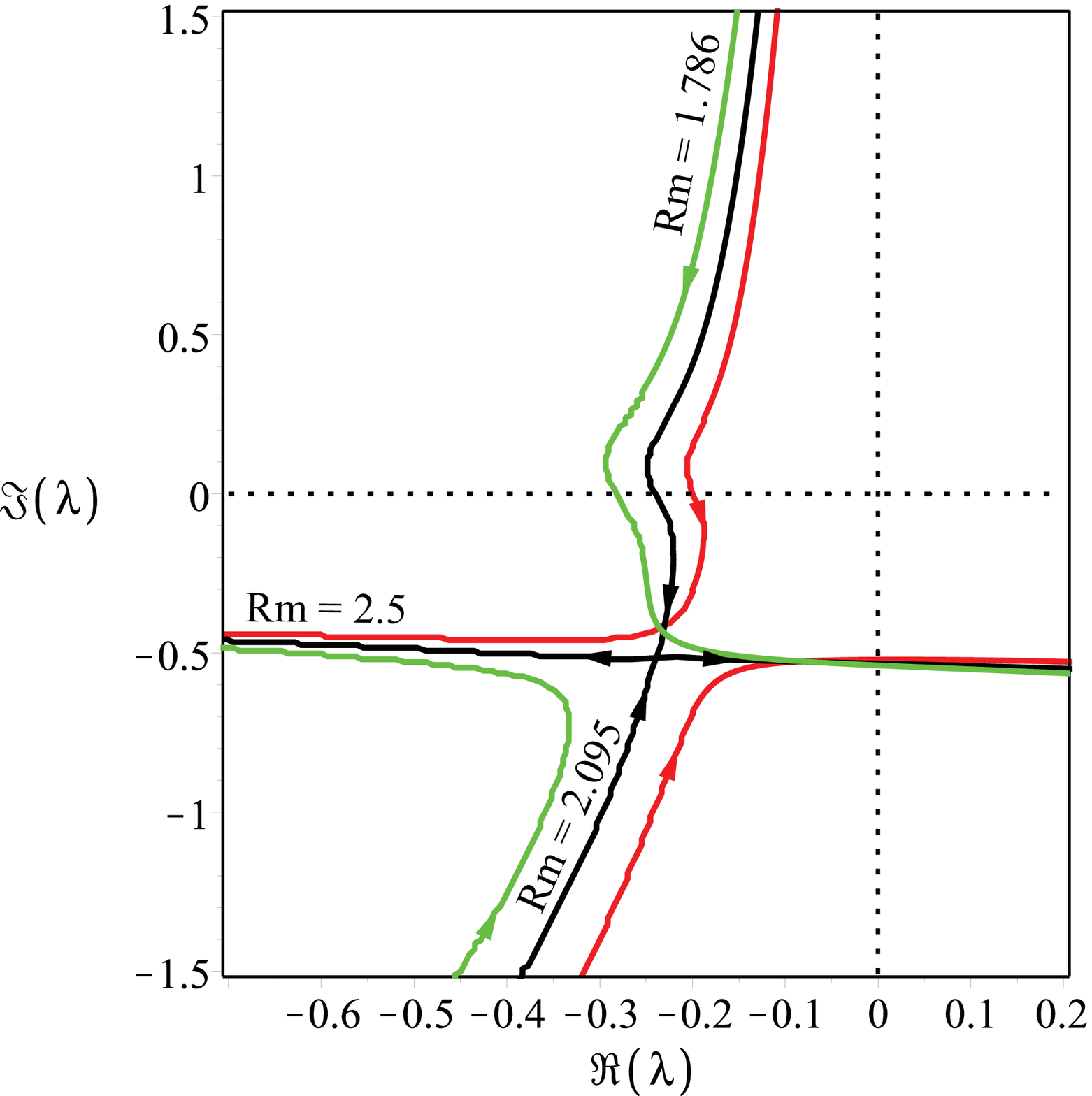}
    \end{center}
    \caption{(Left) Contour plots of the neutral stability surface in the plane of inverse Reynolds numbers at (cuspidal curve) ${\rm Ro}={\rm Ro}_c\approx-1.07855$, (filled area) ${\rm Ro}=-1$, and (intermediate curve) ${\rm Ro}=-1.04$. Two singular Whitney umbrella points (filled diamonds) exist at the intersection of the line ${\rm Pm}=-1$ and the neutral stability curve at ${\rm Ro}=-1$ and another one exists at the origin when ${\rm Ro}={\rm Ro}_c$. From these singularities the lines $\rm EP^{\pm}, EP^0$ of exceptional points are stemming that
    govern the transfer of modes shown in the right panel. (Right) For ${\rm Rb}=-1$, ${\rm S}=1$, $n=\sqrt{2}$, and  ${\rm Re}=1000$ the movement of eigenvalues with decreasing $\rm Ro$ at various ${\rm Rm}$ chosen such that $\rm Pm < 1$. At ${\rm Rm}<1000$ and up to ${\rm Rm}={\rm Rm}_{\rm EP^{-}}\approx 2.095$ it is the branch corresponding to perturbed imaginary eigenvalues with positive Krein sign that causes instability. When ${\rm Rm}={\rm Rm}_{\rm EP^{-}}$ two simple eigenvalues approach each other to merge exactly at $\rm Ro=-1$ into a double eigenvalue whose corresponding matrix is a Jordan block, $\lambda_{\rm EP^{-}}\approx -i0.5086-0.2391$.  At ${\rm Rm}< {\rm Rm}_{\rm EP^{-}}$ the instability shifts to the branch corresponding to perturbed imaginary eigenvalues with negative Krein sign.}
    \label{fig7}
    \end{figure}

Fig.~\ref{fig6}(a) shows that at a fixed ${\rm Pm}\ne 1$ the critical value of $\rm Ro$ at the onset of the double-diffusive AMRI is displaced by an order one distance along the $\rm Ro$-axis with respect to the critical value ${\rm Ro}_c$ of the diffusionless case, when both viscous and ohmic diffusion tend to zero. This effect does not depend on the choice of $n$, Fig.~\ref{fig6}(b). Indeed, the critical values of $\rm Ro$ in the limit of vanishing dissipation at a fixed $\rm Pm$ and ${\rm S}=1$ and ${\rm Rb}=-1$ satisfy the following equation:
\ba{limpm}
({\rm Pm}+1)^3n^2(3{\rm Ro}^2-4n^2-14{\rm Ro}-9)&+&4({\rm Pm}{\rm Ro}-3{\rm Ro}-4)(2{\rm Pm}{\rm Ro}+3{\rm Pm}+1)^2\nn\\
&=&16({\rm Pm}+1){\rm Pm}({\rm Ro}+1)^2n^2.
\ea
Using this equation one can easily check analytically that the critical $\rm Ro$ has its minimum at $\rm Pm=1$ independently of the choice of $n$, Fig.~\ref{fig6}(c), cf. \cite{LD1977}. Nevertheless, the displacement is rather small if ${\rm Pm}\in [0,1]$ with the maximum attained at ${\rm Pm}=0$ where the diffusionless limit of the critical Rossby number is equal to $\frac{5}{3}(2-\sqrt{7})\approx -1.07625<-1$, i.e. weak dissipation with dominating ohmic losses is not capable to destabilize even the Chandrasekhar equipartition solution at ${\rm Ro}=-1$. Does the increase in viscosity and resistivity change this tendency?

\subsubsection{AMRI of the Rayleigh-stable flows at low and high $\rm Pm$ when dissipation is finite}

Indeed, it does. Fig.~\ref{fig6}(d) demonstrates the evolution of the critical Rossby number as a function of ${\rm Rm}^{-1}\in[0,100]$ under the constraint ${\rm Rm}-{\rm Re} {\rm Pm}=0$ at various ${\rm Pm}\in[0,1]$ in the assumption that $\rm Rb=-1$, $\rm S=1$, and $n=\sqrt{2}$. Despite the critical Rossby number does not exceed the value $\rm Ro=-1$ of the equipartition solution for all ${\rm Pm}\in[0,1]$ when ${\rm Rm}^{-1} < 0.1$, it can grow considerably and attain a maximum when ${\rm Rm}^{-1} > 0.1$. For instance, if
${\rm Pm}={\rm Pm}_l\approx 0.0856058$ the maximal critical value is $\rm Ro=-1$, which is attained at ${\rm Rm}={\rm Rm}_l \approx 0.6552421$  (or ${\rm Rm}_l^{-1} \approx 1.5261535$), see Fig.~\ref{fig6}(d) where this maximum is marked by the filled circle. For $0<{\rm Pm}< {\rm Pm}_l$ the maximal critical Rossby number exceeds the value of $\rm Ro=-1$.

In the inductionless limit $(\rm Pm=0)$ the azimutal magnetorotational instability (AMRI) occurs at $\rm Ro\ge -1$ if $\rm Rm \le Rm_*$, where ${\rm Rm_*}=\frac{1}{2}\sqrt{4+2\sqrt{5}}$ (${\rm Rm_*}^{-1}\approx 0.6871$, open circle in Fig.~\ref{fig6}(d)). The critical value of the fluid Rossby number monotonically grows with decreasing $\rm Rm$ attaining its maximal value\footnote{known as the lower Liu limit \cite{Liu2006,KSF2014b,CKH2015}} $\rm Ro^-=2-2\sqrt{2}\approx -0.8284$ at $\rm Rm = 0$.

If $\rm Ro=Rb$, $\rm S=1$, then at $\rm Pm=0$ we have
\ba{rma}
&{\rm Rm}_*^2=\frac{n^2(n^4-12{\rm Rb}^2+16{\rm Rb})-16({\rm Rb}+2)({\rm Rb}^2-n^2)-n((n^2-2{\rm Rb})^2+8({\rm Rb}^2-n^2))\sqrt{n^2+8{\rm Rb}+16}}{32({\rm Rb}^2-n^2)(n^2-{\rm Rb}-2)^2},
&
\ea
in agreement with the results of \cite{KSF2014b}. At $\rm Rb=-1$ and $n=\sqrt{2}$ Eq.~\rf{rma}
yields ${\rm Rm_*}=\frac{1}{2}\sqrt{4+2\sqrt{5}}$.

On the other hand, the lower Liu limit as a function of $n$ and $\rm Rb$ is \cite{KSF2012,KSF2014b}:
\be{lllnrb}
{\rm Ro}^-(n,{\rm Rb})=-2+(n^2-2{\rm Rb})\frac{n^2-2{\rm Rb}-\sqrt{(n^2-2{\rm Rb})^2-4n^2}}{2n^2}.
\ee
Note that ${\rm Ro}^-(n,{\rm Rb})$ attains its maximum $2-2\sqrt{2}$ at $n=\sqrt{-2\rm Rb}$, which explains our choice\footnote{Notice that in the recent paper \cite{SK2015} minimization of the Reynolds and Hartmann numbers over $n$ yielded the critical azimuthal wavenumber $n_c\approx 1.4$, which is close to $n=\sqrt{2}$.} of $n=\sqrt{2}$ for the case when $\rm Rb =-1$, cf. also Fig.~\ref{fig1}(right). Moreover, at $n=\sqrt{-2\rm Rb}$ the instability condition ${\rm Ro} < {\rm Ro^-}$ reduces  to \rf{rorb} after some algebra.

We see that there exists a critical value of the magnetic Prandtl number ${\rm Pm}_l<1$ such that at ${\rm Pm}\in[0, {\rm Pm}_l]$ the Chandrasekhar equipartition solution with $\rm Rb=\rm Ro =-1$, and $\rm S=1$ is destabilized by dissipation when viscosity is sufficiently small and ohmic diffusion is sufficiently large. In contrast, at ${\rm Ro}-{\rm Ro}_c \ll 1$ the marginally stable diffusionless system can be destabilized at $\rm Pm <1$ when both viscosity and resistivity are infinitesimally small, Fig.~\ref{fig4}(left).

In order to understand how these instabilities are related to each other, we plot the neutral stability curves in the plane of inverse Reynolds numbers ${\rm Re}^{-1}, {\rm Rm}^{-1} \in [-0.5,1]$ for ${\rm Ro}\in[{\rm Ro}_c,-1]$, Fig.~\ref{fig7}(left). Although negative Reynolds numbers have no physical meaning, it is instructive to extend the neutral stability curves to the corresponding region of the parameter plane. At ${\rm Ro}={\rm Ro_c}$ the stability domain is inside the area bounded by a curve having a cuspidal singularity at the origin with the tangent line at the cuspidal point specified by the condition $\rm Pm=1$; this geometry yields destabilization by infinitesimally small dissipation at all $\rm Pm \ne 1$.

As soon as $\rm Ro$ departs from ${\rm Ro}_c$, the cusp at the origin transforms into a self-intersection, the angle of which increases with the increase in $\rm Ro$ and becomes equal to $\pi$ at $\rm Ro = -1$. For this reason, at $\rm Ro$ close to $-1$ the neutral stability curve partially belongs to the region of negative Reynolds numbers which makes destabilization by infinitesimally small dissipation impossible for all $\rm Pm>0$. In particular, at $\rm S=1$ and vanishing viscosity the ohmic diffusion is stabilizing in the interval $0<{\rm Rm}^{-1}<{\rm Rm}_*^{-1}$ when
\be{rorm}
{\rm Ro} > {\rm Ro}_{\rm Rm}:=\frac{6{\rm Rb}^2-{\rm Rb}n^2+6n^2-2(n^2-3{\rm Rb})\sqrt{{\rm Rb}^2+3n^2}}{3n^2}.
\ee
At $\rm Rb=-1$ and $n=\sqrt{2}$ we have ${\rm Ro}_{\rm Rm}=\frac{5}{3}(2-\sqrt{7})\approx -1.07625 > {\rm Ro}_c\approx -1.07855$. At $\rm S=1$ and $\rm Ro=Rb$
the critical magnetic Reynolds  number ${\rm Rm}_*$ is defined by Eq.~\rf{rma}.

A similar instability domain exists also in the case of $\rm Pm >1$, Fig.~\ref{fig7}(left). At $\rm Ro=-1$ the ray from the origin with the slope ${\rm Pm}={\rm Pm}_u\approx 11.681451$ is tangent to the boundary of the domain at ${\rm Re}={\rm Re}_u \approx 0.6552421$ (${\rm Re}_u^{-1}\approx 1.5261535$). In particular, in the case of vanishing ohmic dissipation the instability occurs at ${\rm Re}<{\rm Re}_*$ when ${\rm Ro}>{\rm Ro}_{\rm Re}$, where  ${\rm Re}_*$ is given by
\ba{restar}
&{\rm Re}_*^2=\frac{n^6-4({\rm Rb}+1)n^4-4{\rm Rb}^2(3n^2+4{\rm Rb}+8)+n(4({\rm Rb}+2)^2-(n^2-2)^2-12)\sqrt{n^2-8{\rm Rb}}}
{32(({\rm Rb}+2)^2-n^2)(n^2+{\rm Rb})^2}.&
\ea
At $\rm Rb=-1$ and $n=\sqrt{2}$  we have $
{\rm Ro}_{\rm Re}\approx-1.07639$ and ${\rm Re}_*=\frac{1}{2}\sqrt{4+2\sqrt{5}}=\rm Rm_*$.

Hence, the Chandrasekhar equipartition solution $({\rm Ro}={\rm Rb=-1}, {\rm S}=1)$ can be destabilized by dissipation either when $0\le{\rm Pm}<{\rm Pm}_l$ and $0<{\rm Rm}<{\rm Rm}_*$ or when ${\rm Pm}_u<{\rm Pm}<\infty$ and $0<{\rm Re}<{\rm Re}_*$, see Fig.~\ref{fig7}(left) where open circles mark the values of $\rm Re_*$ and $\rm Rm_*$. At $n=\sqrt{2}$  stability of the Chandrasekhar solution is not affected by the double diffusion if ${\rm Pm} \in [0.0856058,11.681451]$.

\subsubsection{Transfer of instability between modes when $\rm Pm$ significantly deviates from 1}

Fig.~\ref{fig7}(left) shows that the neutral stability curves at ${\rm Ro}=-1$ orthogonally intersect the anti-diagonal line with the slope ${\rm Pm}=-1$ at the two exceptional points (marked by the filled diamonds) with the coordinates $({\rm Rm}_{\diamond}^{-1},-{\rm Re}_{\diamond}^{-1})$ and $(-{\rm Rm}_{\diamond}^{-1},{\rm Re}_{\diamond}^{-1})$, where
\be{ephid}
{\rm Rm}_{\diamond}^{-1}={\rm Re}_{\diamond}^{-1}=\frac{\sqrt{2}}{4n}\sqrt{8n^4+20n^2-1-(8n^2+1)^{3/2}}.
\ee
At both exceptional points there exists a pair of simple imaginary eigenvalues and a double imaginary eigenvalue $\lambda_{\diamond}$ with a Jordan block:
\be{epeig}
\lambda_{\diamond}=-i\frac{4 n^2-1-\sqrt{1+8n^2}}{4n}.
\ee
At $n=\sqrt{2}$ Eqs.~\rf{ephid} and \rf{epeig} yield
\be{ephidn}
{\rm Rm}_{\diamond}^{-1}={\rm Re}_{\diamond}^{-1}=\frac{1}{4}\sqrt{71-17\sqrt{17}}\approx0.23811,\quad
\lambda_{\diamond}=-i\frac{7-\sqrt{17}}{4\sqrt{2}}\approx-i0.50857.
\ee
A segment of the anti-diagonal between the exceptional points is a part of the stability boundary at $\rm Ro=-1$ and all the eigenvalues at the points of this segment are imaginary.

            \begin{figure}
    \begin{center}
    \includegraphics[angle=0, width=0.45\textwidth]{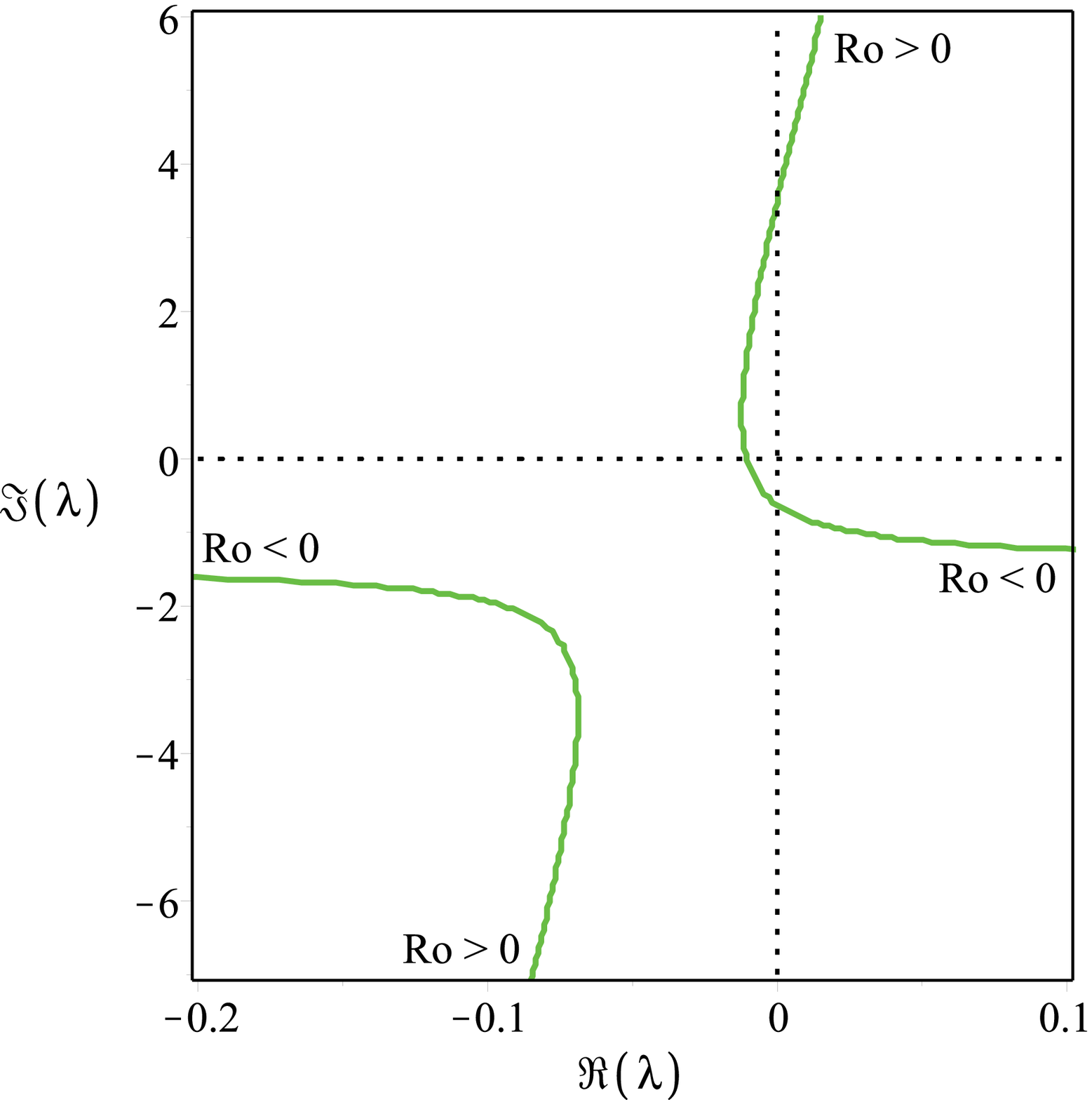}
    \includegraphics[angle=0, width=0.45\textwidth]{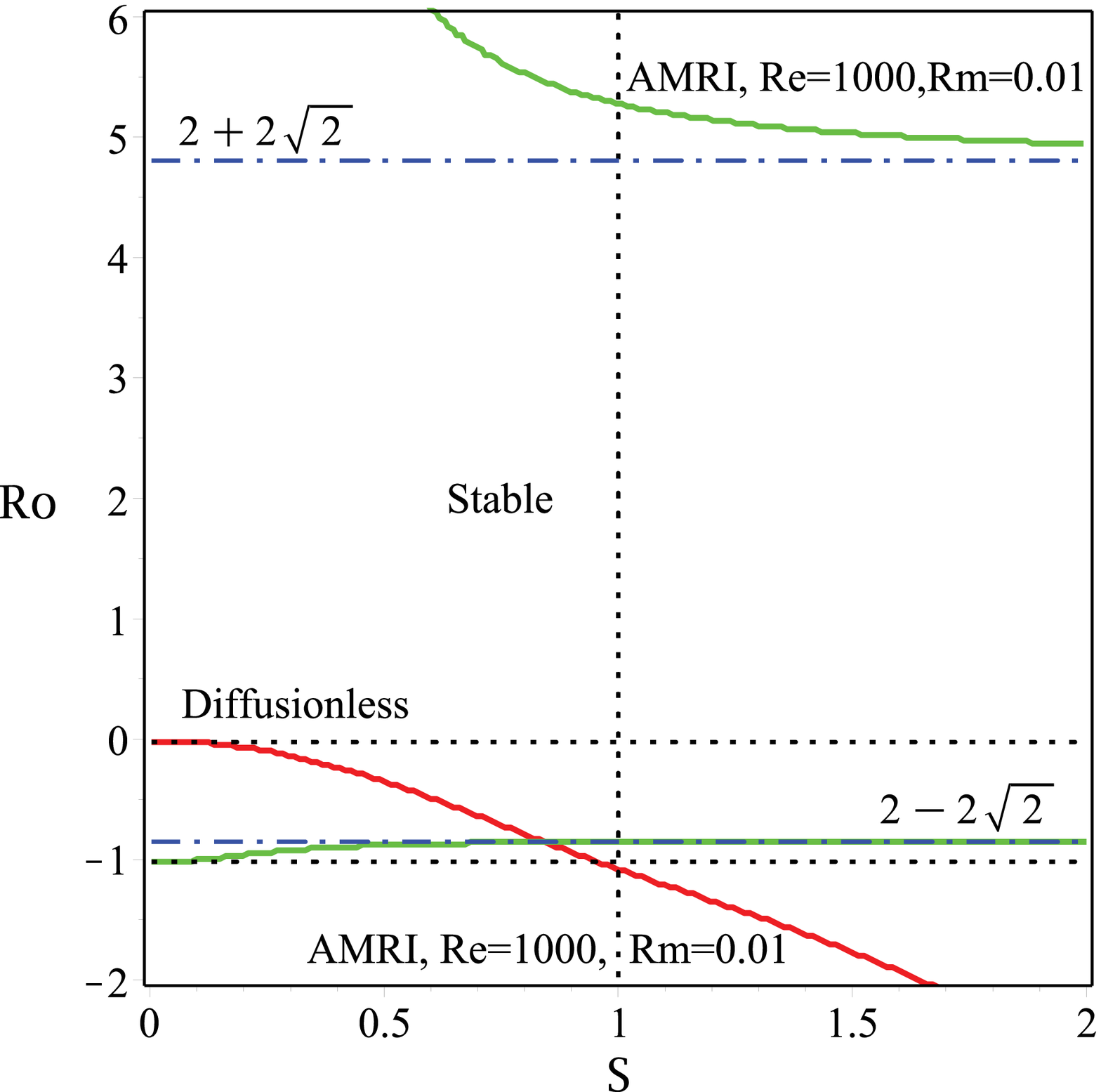}
    \end{center}
    \caption{(Left) For ${\rm Rb}=-1$, $\rm S=1$, and $n=\sqrt{2}$ and fixed $\rm Re=1000$ and $\rm Rm=0.01$ the movement of eigenvalues in the complex plane as $\rm Ro$ is varied demonstrating that at $\rm Pm=10^{-5}$ one and the same eigenvalue branch is responsible for instability both at $\rm Ro <0$ and $\rm Ro >0$. (Right) The corresponding neutral stability curves in the $(\rm S, Ro)$-plane exist below the lower Liu limit of $\rm Ro=2-2\sqrt{2}$ (destabilizing the Chandrasekhar equipartition solution) and above the upper Liu limit of $\rm Ro=2+2\sqrt{2}$ that are attainable only at $\rm Re \rightarrow \infty$ and $\rm Rm \rightarrow 0$. In contrast, the diffusionless AMRI exists above the lower Liu limit at small $\rm S$ but does not affect the Chandrasekhar equipartition solution at $\rm S=1$.}
    \label{fig8}
    \end{figure}

We see that the domain of asymptotic stability at $\rm Ro=-1$ extends to the region of negative Reynolds numbers and that at the constraint $\rm Rm=-\rm Re$ the double-diffusive system has imaginary spectrum on the interval between the two exceptional points. If we interpret the negative dissipation as an energy gain, then, formally, we could say that at $\rm Rm=-\rm Re$ the energy gain is compensated by the energy loss. Non-Hermitian systems in which gain and loss are balanced are known as parity-time (PT) symmetric systems \cite{K2013pt,Kottos2011}. The interval of marginal stability of the PT-symmetric system forms a self-intersection singularity on the stability boundary of a general dissipative system with the Whitney umbrella singularities at the exceptional points corresponding to double imaginary eigenvalues \cite{K2013pt,K2013dg}.
Therefore, the neutral stability surface of our double-diffusive system contains the interval of self-intersection on the $\rm Ro$-axis (${\rm Ro} > {\rm Ro}_c$) that is orthogonal at $\rm Ro=-1$ to the interval of the anti-diagonal with the slope ${\rm Pm}=-1$ confined between the two exceptional points. At the exceptional points of this interval and at the exceptional point on the $\rm Ro$-axis at ${\rm Ro}={\rm Ro}_c$ the neutral stability surface in the $({\rm Rm}^{-1},{\rm Re}^{-1},{\rm Ro})$-space has three Whitney umbrella singularities. The singularities `hidden' in the region of negative Reynolds numbers are responsible for the separation of domains of AMRI due to weak or strong dissipation.

It turns out, that this separation is not only quantitative but also qualitative, as comparison of the movement of eigenvalues demonstrates at fixed $\rm Re =1000$ and $\rm Rm=500$ in Fig.~\ref{fig5}(left)
and at $\rm Re =1000$ and $\rm Rm\approx 1.789$ in Fig.~\ref{fig7}(right). In both cases $\rm Pm < 1$. However, in the case of $\rm Pm=0.5$ it is the branch with lower negative frequencies corresponding to the perturbed imaginary eigenvalues with positive Krein sign of the diffusionless Hamiltonian system that becomes unstable due to prevailing ohmic diffusion. In contrast, at much smaller $\rm Pm \approx 0.001789$ the instability moves to a branch with  higher negative frequencies that can be seen as stemming from the imaginary eigenvalues with negative Krein sign of the diffusionless Hamiltonian system. Keeping $\rm Re =1000$ and slightly increasing  the magnetic Reynolds number to $\rm Rm \approx 2.095$ we see at $\rm Ro=-1$ the crossing of the eigenvalue branches at the double eigenvalue $\lambda_{EP^-}\approx -i0.5086-0.2391$. The crossing transforms into another avoided crossing when $\rm Rm = 2.5$. At $\rm Rm = 2.5$, again, it is the branch corresponding to higher negative frequencies (positive Krein sign) that is destabilized by dissipation, Fig.~\ref{fig7}(right).

In fact, when $\rm Re =1000$ is given, the branch corresponding to the unperturbed imaginary eigenvalues with  positive Krein sign is destabilized by dissipation when the magnetic Reynolds number decreases from $\rm Rm=1000$ ($\rm Pm=1$) to $\rm Rm \approx 2.095$ ($\rm Pm\approx 0.002095$). As soon as $\rm Rm < 2.095$ ($\rm Pm< 0.002095$) the instability is transferred to a branch corresponding to the unperturbed imaginary eigenvalues with negative Krein sign.
The reason is the existence of a set in the stability domain corresponding to double complex eigenvalues. This set exists at $\rm Ro=-1$ and consists of the two straight lines
\be{epl}
\frac{1}{\rm Re}=\pm\frac{2}{{\rm Re}_{\diamond}}+\frac{1}{\rm Rm}
\ee
that are tangent to the neutral stability curves at the exceptional points with the coordinates $({\rm Rm}_{\diamond}^{-1},-{\rm Re}_{\diamond}^{-1})$ and $(-{\rm Rm}_{\diamond}^{-1},{\rm Re}_{\diamond}^{-1})$, where
${\rm Rm}_{\diamond}$ and ${\rm Re}_{\diamond}$ are defined by Eq.~\rf{ephid}.

In Fig.~\ref{fig7}(left) the lines corresponding to different signs in Eq.~\rf{epl} are marked as ${\rm EP}^+$ (the upper dot line) and ${\rm EP}^-$ (the lower dot line). At the points of the $\rm EP$-lines \rf{epl} there exist double complex eigenvalues (exceptional points) $\lambda_{\rm EP^{\pm}}$ given  by the expression
\be{dep}
\lambda_{\rm EP^{\pm}}=\lambda_{\diamond}-\left(\frac{1}{\rm Rm}\pm\frac{1}{{\rm Rm}_{\diamond}} \right).
\ee
At $n=\sqrt{2}$ and ${\rm Re}_{\rm EP^-}=10^3$, we find that $\frac{1}{\rm Rm_{\rm EP^-}}=\frac{2}{\rm Re_{\diamond}}+\frac{1}{\rm Re_{\rm EP^-}}\approx0.477$ ($\rm Rm_{\rm EP^-}\approx 2.095$) and
\be{laepn}
\lambda_{\rm EP^{-}}=i\frac{\sqrt{17}-7}{4\sqrt{2}}-\frac{1}{4}\sqrt{71-17\sqrt{17}}-\frac{1}{\rm Re_{\rm EP^-}}\approx-0.2391-i0.5086.
\ee

We see that the three Whitney umbrella points and, related to them, three lines of double complex eigenvalues (marked in Fig.~\ref{fig7}(left) as $\rm EP^{\pm}$ and $\rm EP^{0}$) actually control the dissipation-induced destabilization acting as switches of unstable modes. The singular geometry of the neutral stability surface guides the limiting scenarios and connection of the double-diffusive system to a Hamiltonian or to a PT-symmetric one.

\subsubsection{Connection between the lower and upper Liu limits at $\rm Pm \ll 1$}

Let us keep $\rm Re=1000$ and allow the magnetic Reynolds number to decrease beyond the critical value $\rm Rm_{\rm EP^-}\approx 2.095$. During this process the pattern of interacting eigenvalues remains qualitatively the same, cf. Fig.~\ref{fig7}(right) and Fig.~\ref{fig8}(left). However, an important new feature appears as the magnetic Prandtl number approaches the inductionless limit $\rm Pm=0$. Indeed, at $\rm Re=1000$ and $\rm Rm=0.01$ corresponding to $\rm Pm=10^{-5}$ one and the same eigenvalue branch has unstable parts both at $\rm Ro <0$ and at $\rm Ro >0$, see Fig.~\ref{fig8}(left). This is in striking contrast to the case of moderately small magnetic Prandtl numbers shown in Fig.~\ref{fig7}(right) or to the diffusionless case when the instability occurs only at $\rm Ro <0$.

The Bilharz criterion reveals two regions of instability in the $(\rm S, Ro)$-plane for $\rm Rb=-1$, $n=\sqrt{2}$ and $\rm Re=1000$ and $\rm Rm= 0.01$, Fig.~\ref{fig8}(right). The first one exists at $\rm Ro < 2-2\sqrt{2}<0$ and the second one at $\rm Ro > 2+2\sqrt{2}>0$. In the gap between the lower Liu limit $(2-2\sqrt{2})$ and the upper Liu limit $(2+2\sqrt{2})$ the system is stable \cite{KSF2014b,Liu2006}. Both Liu limits are attained when $\rm Re \rightarrow \infty$ and $\rm Rm \rightarrow 0$. If the double-diffusive instability domain at $\rm Ro<0$ can be considered as a deformation of the instability domain of the diffusionless system, the instability of the magnetized circular Couette-Taylor flow in super rotation \cite{SK2015} at $\rm Ro>0$ turns out to exist only in the presence of dissipation. Remarkably, the two seemingly different instabilities are caused by the eigenvalues living on a single eigenvalue branch in the complex plane, Fig.~\ref{fig8}(left).

The oscillatory instability at $\rm Pm \ll 1$ of a circular Couette-Taylor flow in an azimuthal magnetic field with $\rm Rb=-1$ and $\rm Ro < 2-2\sqrt{2}$, i.e. the azimuthal magnetorotational instability (AMRI), has already been observed in recent experiments with liquid metals \cite{S2014}. We therefore identify the observed \textit{inductionless AMRI at ${\rm Pm}\ll 1$ as simply the manifestation of a dissipation-induced instability of waves of negative energy of the diffusionless system caused by the prevailing ohmic diffusion}. In particular, at $\rm Ro=Rb=-1$ and $\rm S=1$ the inductionless AMRI is the dissipation-induced instability of the Chandrasekhar equipartition solution.

\section{Conclusion}

We have studied azimuthal magnetorotational instability (AMRI) of a circular Couette-Taylor flow of an incompressible electrically conducting Newtonian fluid in the presence of an azimuthal magnetic field of arbitrary radial dependence. With the use of geometrical optics asymptotic solutions we have reduced the problem to the analysis of the dispersion relation of the transport equation for the amplitude of a localized perturbation. We have represented the corresponding matrix eigenvalue problem in the form of a Hamiltonian diffusionless system perturbed by ohmic diffusion and fluid viscosity. We have established that the diffusionless AMRI corresponds to the Krein collision of simple imaginary eigenvalues with the opposite Krein (or energy) sign and have derived an analytic expression for the instability threshold of the diffusionless system using the discriminant of the complex polynomial dispersion relation. We have demonstrated that the threshold of the double-diffusive AMRI with equal viscosity and electrical resistivity $(\rm Pm=1)$ smoothly converges to the threshold of the diffusionless AMRI in the limit of the infinitesimally small dissipation and this result does not change when other parameters are varied.

In contrast with the case when the coefficients of viscosity and resistivity are equal, the prevalence of resistivity over viscosity or vice-versa causes the azimuthal magnetorotational instability in the parameter regions where the diffusionless AMRI is prohibited, for instance, in the case of super-rotating flows. In particular, non-equal and finite viscosity and resistivity destabilize the celebrated Chandrasekhar energy equipartition solution. Analyzing the neutral stability surface of the double-diffusive system we have found that:
\begin{itemize}
\item marginally stable Hamiltonian equilibria of the diffusionless system form an edge on the neutral stability surface of the double-diffusive system that ends up with the Whitney umbrella singular point at the onset of the Hamilton-Hopf bifurcation;
\item another edge with the two Whitney umbrella singular points at its ends corresponds to marginally stable double-diffusive systems with the balanced energy gain and loss (PT-symmetric systems);
\item three codimension-2 sets corresponding to complex double-degenerate eigenvalues with Jordan blocks (exceptional points) stem from each of the Whitney umbrella singularities and live in the stability domain of the double-diffusive system;
\item the sets of exceptional points control transfer of instability between modes of positive and negative energy whereas the Whitney umbrellas govern the limiting scenarios for the instability thresholds including the case of vanishing dissipation;
\item AMRI can be interpreted as an instability of the Chandrasekhar equipartition solution induced by finite dissipation when  either  ${\rm Pm}\in [0, 1)$ is sufficiently small or ${\rm Pm}\in (1, \infty)$ is sufficiently large.
\item inductionless AMRI occurring both at $\rm Ro< 0$ and $\rm Ro >0$ when $\rm Pm \ll 1$ is caused by the eigenvalues of the one and the same branch stemming from the negative energy modes of the diffusionless system, as in the classical dissipation-induced instability.
\end{itemize}

\vskip6pt

The author is thankful for
partial support through the EU FP7 ERC grant ERC-2013-
ADG-340561-INSTABILITIES.

\end{document}